\documentclass[12pt,notitlepage]{article} 

\usepackage{amsmath}
\usepackage{amssymb}
\usepackage{latexsym}
\usepackage{amsfonts}
\usepackage{endnotes} 
\usepackage{multirow}
\usepackage{graphicx}
\usepackage{float}
\usepackage[margin=2.5cm]{geometry}

\newcommand{\nn}{\nonumber\\}

\newcommand{\tK}{{\tilde\K}}
\newcommand{\tE}{{\tilde\E}}
\newcommand{\EK}{\frac{\mathbb E}{\mathbb K}}
\newcommand{\tk}{{\tilde k}}
\newcommand{\td}{{\tilde d}}
\newcommand{\Ld}{\Lambda}
\newcommand{\E}{\mathbb E}
\newcommand{\K} {\mathbb K}
\newcommand{\Sp} {\mathcal S}
\newcommand{\op}{\ord}
\newcommand{\dd}{\partial^2_\sigma}
\newcommand{\bea}{\begin{eqnarray}}
\newcommand{\ea}{\end{eqnarray}}
\newcommand{\ord}{\,{\cal O}}

\begin{document}

\title{A semiclassical analysis of the fluctuation eigenvalues and the one-loop energy of the folded spinning superstring in $AdS_5 \times S^5$ } 

\author{G\"ok\c ce Ba\c sar \\ {\it Department of Physics, University of Connecticut, Storrs CT 06269, USA} }
	
\date{}

\maketitle

\begin{abstract}
We systematically construct a semiclassical expansion for the eigenvalues of the $2^{nd}$ order quantum fluctuations of the folded spinning superstring rotating in the $AdS_3$ part of $AdS_5 \times S^5$ with two alternative methods; by using the exact expression of the Bloch momentum generated by the curvature induced periodic potentials and by using the large energy expansion of the dispersion relation. We then calculate the one-loop correction to the energy by summing over the eigenvalues. Our results are extremely accurate for strings whose ends are not too close to the AdS radius. Finally we derive the small spin Regge expansion in the context of zeta function approximation.  
\end{abstract}

\section{Introduction}

The duality between type IIB superstring theory in $AdS_5 \times S^5$ and $\mathcal N$=4 Super-Yang Mills theory in the planar limit \cite{Maldacena} has opened a new window in understanding gauge theories in strong coupling limit. The identification $ R^4/\alpha^{\prime\,2}\equiv
\lambda=g_{YM}^2N$, relates the small string tension limit of the string theory to the strong 't Hooft coupling limit of the gauge theory. In particular, single 
string states of the free string theory are related to the single trace gauge invariant operators of planar SYM. Until recently, analytical calculations, typically used to be restricted either to the classical supergravity limit ($\lambda \gg 1$) or to the perturbative gauge theory limit ($\lambda \ll 1$). However, in general, the observables of the gauge theory are nontrivial functions of $\lambda$ unless they are protected by supersymmetry such as BPS states and one needs to interpolate between strong and weak 't Hooft coupling regimes to check the theory. Therefore, going beyond the supergravity approximation was a crucial step on this path and extended the domain of analytical calculations vastly.  

One important particular manifestation of the duality is the identification of the energy (defined with respect to the global time coordinate) of a single string with the scaling dimension of the corresponding single trace operator. The spins of the string in $AdS_5$ and $S^5$ sectors translate into charges of the operators under $SO(4,2)$ (conformal) and $SU(4)$ (R) symmetries of the SYM respectively. Strings with large spins has been analyzed by semiclassical methods \cite{Gubser:2002tv,Berenstein:2002jq,deVega:1996mv,Frolov:2002av} and the underlying integrability property of the theory by constructing an algebraic curve to obtain Bethe type equations \cite{algcurve1,algcurve2,SchaferNameki:2005is,integ}. The corresponding scaling dimensions of SYM are also calculated perturbatively up to 5 loops \cite{4loop} and using Bethe ansatz techniques where one can map the problem into a spin chain system and obtain the thermodynamic limit of the Bethe equations \cite{novel,Gromov}. Therefore both explicit solutions and equations that govern the energy/scaling dimension are compared \cite{integ,novel,dorey,spiky,bethe,belitsky}. The small spin limit is also investigated both in string side \cite{Frolov:2002av,short} in the semiclassical perspective and in the gauge side \cite{Gromov} by extending the integrability methods to short operators.

Here we consider a folded string spinning in the AdS part of $AdS_5\times S^5$ with a single Lorentz spin $S$ \cite{Gubser:2002tv,deVega:1996mv,Frolov:2002av}. In the dual gauge theory, this setup corresponds to a gauge invariant, single trace, twist 2 operator. The states with small spin, 
describe short operators such as the Konishi operator tr$(\Phi^2)$ and the states with large spin describe long operators like tr$(\Phi D_+^S \Phi)$ with $S\gg1$. 
Classically, the string state is a soliton and its energy obeys the ordinary Regge relation $E=\lambda^{1/4}\sqrt{2S}+...$ for small $S$. In the opposite regime, 
where the spin is large, one finds a logarithmic behavior $E= S+\frac{\sqrt{\lambda}}{\pi}$ln$S$+... The same logarithmic scaling has been also found in 
weakly coupled SYM, with a coefficient analytic in $\lambda$, which suggests the strong and weakly coupled regimes are connected by a smooth 
function at large $S$ \cite{Benna:2006nd}. This function is known as the \textit{cusp anomalous dimension} \cite{cusp,Basso:2007wd} and governs the anomalous dimension of a light like Wilson-loop with a cusp. It also extends to certain deep inelastic scattering properties of QCD \cite{QCD}. When the spin is small, the relation between the weak and strong coupling regimes is more subtle and unlike the large spin case, the small spin limit and the strong/weak coupling limits do not seem to commute \cite{short}. 

The one-loop $\frac{R^2}{\alpha^\prime}=\frac{1}{\sqrt{\lambda}}$ corrections to the classical energy are also calculated in the semiclassical framework \cite{Frolov:2002av,short,long}, considering the small and large spin limits are separately. Then, it has been recently shown \cite{Beccaria:2010ry} that, in the static gauge, the  $2^{nd}$ order fluctuations of different modes decouple in the string sigma model. Furthermore, the curvature terms introduce periodic kink-like potentials which have a special form, namely the single-gap Lam\'e form \cite{whittaker,algebro}, and the associated fluctuation equations are analytically soluble. This leads to simple explicit integral representation expressions for the one loop energy corrections, for an arbitrary spin, even though there is no simple closed-form expression for the fluctuation eigenvalues. These explicit one-loop corrections have been studied in the small and large spin limits \cite{Beccaria:2010ry}.

Our goal in this paper is to study the calculation of the one-loop energy correction by summing over the fluctuation eigenvalues, providing a complementary perspective on the strategy used in \cite{Beccaria:2010ry}. We first find an accurate expansion for the fluctuation eigenvalues by constructing a WKB series. The starting point is the general theorem \cite{magnus} that states 
that in the large $n$ limit,  the eigenvalues of the equation -$\phi_n^{\prime\prime} +V(x)\phi_n=\lambda_n\phi_n$ satisfying the properties $V(x)=V(x+L)$, $\langle V \rangle$=0 
and periodic boundary conditions can be expanded as:     
\bea
\Lambda_{2n-1}=(2n)^2+\frac{\langle V^2 \rangle}{(2n)^2}+\ord(n^{-2}) \nn
\Lambda_{2n}=(2n)^2+\frac{\langle V^2 \rangle}{(2n)^2}+\ord(n^{-2})
\label{magnus}
\ea
where $\langle V^k\rangle\equiv\frac{1}{L}\int dx\, (V(x))^k$. Further, we extend this expansion for $\langle V \rangle \neq 0$ and develop two methods to generate terms up to  any desired order $\ord(n^{-2m})$; 
first by using the closed-form expression for the Bloch momentum in these soluble Lam\'e systems and second, by using the large energy expansion of the 
resolvent. (This second method can be used for other periodic, but not necessarily finite-gap, fluctuation potentials.) The expansion gives surprisingly accurate results even for the lowest eigenstates and is applicable for all spins lengths except the very long string regime where string extends to the AdS boundary (the large spin regime). In this manner, it can be realized as a complementary result to the large 
spin result $\Lambda_n=n^2+C\kappa^2$ of \cite{Frolov:2002av}. We will then use this expansion to calculate the one-loop energy in various ways. 
 
The organization of the paper is as follows:

We will begin with a brief recapitulation of the classical soliton solution of the spinning string in $AdS_5 \times S^5$ \cite{Gubser:2002tv, Frolov:2002av} 
and its leading Regge trajectories in the small and large spin regimes in section \ref{classical}. 
The Regge relation between the classical energy and spin is also given implicitly through the elliptic parameter.  
In section \ref{fluctuations} we  introduce the quantum fluctuations and summarize the recently found analytical solutions \cite{Beccaria:2010ry}. Then we construct the expansion for the fluctuation eigenvalues as described above and discuss the applicability of our approximations in 
detail.After representing the eigenvalues as WKB series, we then calculate the one-loop energy by summing over the eigenvalues. Again two different 
approaches will be taken. In the first one we  approximate the sum by an integral using the Euler-Maclaurin formula and in the second one by Riemann zeta functions. The final expression is expressed in terms of the elliptic parameter and we discuss the validity of 
the expansions and show that it works extremely well for both small and intermediate spins, but breaks down in the very long string regime. Hence, our results analytically interpolate between the small spin/short string and large spin/long string limits. We also compare our results with the short spin expansion derived in \cite{Beccaria:2010ry}. Finally in the last section, the one-loop correction to the Regge relation in the small spin limit is discussed in the zeta function context and we show that they agree with the previously obtained results \cite{Frolov:2002av,Beccaria:2010ry}, using a combination of analytic and numerical results.

\section{Classical equations of motion and Regge relations}\label{classical}

The folded spinning string in $AdS_5 \times S^5$:
\bea
ds^2&=&R^2ds^2(AdS_5)+R^2ds^2(S^5) \nn 
ds^2(AdS_5)&=& -\cosh^2 \rho\,dt^2+d\rho^2+\sinh^2\rho\left(d\beta_1^2 +\cos^2\beta_1\,d\beta_2^2+\cos^2\beta_1\cos^2\beta_2\,d\phi^2 \right)
+ds^2(S^5) \nn 
ds^2(S^5)&=&d\theta_1^2 +\cos^2\theta_1(d\theta_2^2+\cos^2\theta_2\ (d\theta_3^2+\cos^2\theta_3(d\theta_4^2+\cos^2\theta_4\,d\theta_5)))
\ea
which is rotating in the $AdS_3$ subspace, can be described by the following ansatz \cite{Gubser:2002tv, deVega:1996mv, Frolov:2002av}:
\bea
t=\kappa\tau, \quad \phi=\omega\tau, \quad \rho=\rho(\sigma), \quad \beta_i=\theta_j=0 \quad (i=1,2\,\,;\,\,j=1,..,5) 
\ea
where $\tau$ and $\sigma$ are the string world-sheet coordinates and $\sigma$ is identified with $\sigma+2\pi$. The equation of motion in the conformal 
gauge $g^{ab}\sqrt{-g}=h^{ab}\sqrt{-h}=$diag(-1,1) reads\footnote{Here $h_{ab}=G_{\mu\nu}\partial_aX^{\mu}\partial_bX^{\nu}$ where $G_{\mu\nu}$ is the 10-d 
target space metric and $g_{ab}$ is the world-sheet metric.}:
\begin{equation}
\rho^{\prime\,2}=\kappa^2\cosh^2\rho-\omega^2\sinh^2\rho
\label{e_mot}
\end{equation}
where $\prime\equiv\frac{d}{d\sigma}$. The general solution of (\ref{e_mot}) can be expressed in terms of the Jacobi elliptic function:
\bea
\rho^\prime(\sigma)=\kappa\,{\rm sn}(\omega \sigma+C;k^2)
\label{rhop}
\ea
with the elliptic parameter $k=\frac{\kappa}{\omega}$ ranging from 0 to 1. We choose to parametrize the string such that $\sigma=0$ coincides with the 
center $\rho=0$. This choice\footnote{In the rest of the paper $\K=\K(k^2)\equiv\int_0^{\pi/2}(1-
k^2\,sin^2\theta)^{-1/2}d\theta$, and $\E=\E(k^2)\equiv\int_0^{\pi/2}(1-k^2\,sin^2\theta)^{1/2}d\theta$ denote the complete elliptic integrals of the first and 
second kind, respectively \cite{whittaker,abram}. We also use the prime notation: $\K^\prime(k^2)\equiv\K(1-k^2)$, $\E^\prime(k^2)\equiv\E(1-k^2)$.} fixes the integration constant C=$\K(k^2)$. In particular the sn function 
is periodic with the period $4\K$ which determines $\omega=\frac{2\K}{\pi}$.  

The end points of the string correspond to the points where $\rho^\prime=0$ or equivalently $\sigma=\frac{\pi}{2},\,\frac{3\pi}{2}$. These points also 
determine the length of the string in units of AdS radius $\rho$ where the string extends to its maximum value $\rho_0$;
\bea
\rho_0=\rho(\frac{\pi}{2})={\rm ln}(\frac{1+k}{\sqrt{1-k^2}} )
\ea
As a result the solution (\ref{rhop}) describes a string that extends from 0 to $\rho_0$ and folds at the turning points $\sigma=\frac{\pi}{2},\,\frac{3\pi}{2}$. 
Around $k=0$, the string is short and spins around the center of AdS where the curvature is small. This corresponds to the flat-space limit of the theory. 
Whereas as $k$ gets close to 1, the end point $\rho_0$ approaches to the boundary of AdS.    

The classical energy and spin are the conserved momenta under $t$ translations and $\phi$ rotations:
\bea
E_0&=&-\frac{1}{2 \pi \alpha^{\prime}}\int_0^{2\pi}\,d\sigma G_{tt} \partial_\tau t=\frac{\kappa R^2}{2 \pi \alpha^{\prime}}\int_0^{2\pi}\,d\sigma \cosh^2\rho=
\frac{2\sqrt{\lambda}}{\pi} \frac{k}{(1-k^2)}\E
\label{classical_energy}
\ea
\bea 
S&=&\frac{1}{2 \pi \alpha^{\prime}}\int_0^{2\pi}\,d\sigma G_{\phi\phi} \partial_\tau \phi=\frac{\omega R^2}{2 \pi \alpha^{\prime}}\int_0^{2\pi}\,d\sigma 
\sinh^2\rho=\frac{2\sqrt{\lambda}}{\pi}\left(\frac{1}{1-k^2}\E-\K \right)
\label{classical_spin}
\ea
At the last step we identified $R^2/\alpha^{\prime}$ with $\sqrt{\lambda}$. Notice that the spin is a monotonously increasing function of $k$ as expected which 
means the spin of the string increases with its length.  
After expressing the classical energy and spin in terms of $k^2$, we look for the Regge relations. In order to write the energy as a function of spin, one 
needs to invert (\ref{classical_spin}) and plug it into (\ref{classical_energy}). However this does not seem possible in general. Instead we expand 
(\ref{classical_spin}) and (\ref{classical_energy}) around small/large spin regimes and write down the Regge relations as expansions. It is convenient to 
work with the "dimensionless" energy and spin ${\mathcal E_0}=\sqrt{\lambda}E_0,\,\Sp=\sqrt{\lambda}S$ where the spin is measured in units of $\frac{R}{\sqrt{\alpha^\prime}}$ and it is explicit that small $S$ implies that $\Sp$ is small and R is large at the same time (or vice-versa). 
In the small spin regime, $\Sp$ has the following expansion;
\bea
\Sp\approx\frac{k^2}{2}+\frac{9 k^4}{16}+\frac{75 k^6}{128}+\ord(k^8)
\ea
After inverting the series and substituting into the energy and expanding it we obtain;
\bea
\mathcal E_0\approx\sqrt{2 \Sp}\left(1+\frac{3}{8}\Sp-\frac{21}{128}\Sp^2+\ord(\Sp^3)\right) 
\ea
Here the first term is the ordinary flat space Regge term.
On the other hand, in the opposite limit where spin is large;
\bea
\Sp\approx\frac{2}{k^2 \pi }+\frac{-1-4 \text{ln}2+\text{ln}\left(k^2\right)}{2 \pi }+\frac{k^2 \left(3-8 \text{ln}2+2 \text{ln}\left(k^2\right)\right)}{32 \pi }+...
\ea
that implies;
\bea
\mathcal E_0\approx\Sp+\frac{{\rm ln}(8\pi\Sp)-1}{\pi}+\frac{{\rm ln}(8\pi\Sp)-1}{2\pi^2\Sp}+...
\ea
Here the coefficient of the leading ln term is recognized as the cusp anomalous dimension in the strongly coupled gauge theory side.

\section{Fluctuation operators and their eigenvalues}
\label{fluctuations}

In order to calculate the one-loop energy correction, we take the quantum fluctuations of the coordinates into account and expand them up to second 
order in the action. In the conformal gauge, the fluctuations of the coordinates $t$,$\rho$ and $\phi$ are coupled \cite{Frolov:2002av}. Neither their solutions nor their 
determinants (which are the main objects of the one-loop energy) seem to be expressible in closed forms. However passing to the static gauge where 
only the coordinates perpendicular to the embedded world-sheet are allowed to fluctuate: 
\bea
{\tilde t}={\tilde \rho}=0
\ea
simplifies the expressions vastly. In the static gauge all the fluctuations are decoupled; and can be computed separately as they factor out in the path 
integral. At  first sight working in the static gauge seems problematic because the $\phi$ mass has a divergent piece and one may be think that it 
compromises the UV finiteness of the theory. But this is not an actual physical problem, rather an artifact of the static gauge \cite{Frolov:2002av,Beccaria:2010ry}. Indeed, the singularity does not have any effect on either the determinants of the fluctuation operators or their eigenvalues. The reason is that, the 
eigenvalues (and the determinants) are only sensitive to the behavior of the potential under a shift by a period and determined basically by the boundary 
conditions. Also in \cite{Beccaria:2010ry} the equality between the static and conformal gauge determinants has been showed explicitly by comparing the (exact) numerical determinants calculated in the conformal gauge with the analytical expressions obtained the static gauge. 

The bosonic part of the fluctuation action in the static gauge is composed of the fluctuations of the remaining 3 of the $AdS_5$ coordinates $\tilde\phi$, $
\tilde\beta_1$,$\tilde\beta_2$ and the 5 coordinates $S^5$; $\tilde\theta_i$. The AdS fluctuations acquire $\sigma$ dependent masses due to curvature 
terms whereas the $S^5$ fluctuations are massless. We also scale $\tilde\phi$ and $\tilde\beta_i$ such that the kinetic terms do not have any extra prefactor. As 
a result we obtain the basic form;
\bea
{\tilde S_B}=-\frac{\sqrt{\lambda}}{4 \pi}\int d^2\sigma(\partial_a{\tilde \phi}\partial^a{\tilde \phi}+m_\phi^2{\tilde \phi}^2+\partial_a{\tilde \beta_i}
\partial^a{\tilde \beta_i}+m_\beta^2{\tilde \beta_i}^2+\partial_a{\tilde \theta_i}\partial^a{\tilde \theta_i})
\ea
where the space dependent masses have simple forms represented in Jacobian elliptic functions: 
\bea
m_\phi^2&\equiv&V_\phi= 2\rho^{\prime 2}+\frac{2\omega^2 \kappa^2}{\rho^{\prime 2}}=2 \kappa^2\,{\rm sn}^2(\omega\sigma+\K;k^2)+2 \omega^2\,{\rm 
ns}^2(\omega\sigma+\K;k^2)\nn
m_\beta^2&\equiv& V_\beta= 2 \rho^{\prime}=2 \kappa^2\,{\rm sn}^2(\omega\sigma+\K;k^2)
\ea
 
The fermionic fluctuations have 8 degrees of freedom and can be organized in 4 $\oplus$ 4 copies of 2d Majorana spinors \cite{Frolov:2002av}.  
In the chiral basis, where $\gamma^3={\rm diag}(I_{4\times4},-I_{4\times4})$, the action reads:
\bea
{\tilde S_F}=-\frac{\sqrt{\lambda}}{4 \pi}\int d^2\sigma 2i ({\bar \psi}\gamma^a\partial_a \psi-im_\psi\bar{\psi}\gamma^3 \psi)
\ea
with the mass term $m_\psi=\rho^\prime$. It is more convenient to square the Dirac operator and work with the $2^{nd}$ order fluctuation operators:
\bea
\op_{\psi\pm}=\partial_\tau^2-\dd+m_{\psi\pm}^2
\ea

The resulting operators have space dependent masses $m_{\psi\pm}^2=m_\psi^2\pm m_\psi^{\prime}=\rho^{\prime\,2}\pm \rho^{\prime\prime\,2}$. These 
type of potentials are "superpartners" in the context of quantum mechanical supersymmetry and they are always isospectral except the zero modes. Their 
spectrum may differ by a zero mode due to the existence of non-normalizable zero modes. However, in the present case, the potentials are periodic, and both of them have non singular zero modes. In fact, $m_{\psi\pm}^2$ are the same function, shifted by half a period with respect to one another; they are "self-isospectral" \cite{Dunne:1997ia} and  have identical spectra, including the zero modes. In terms of the Jacobi functions they are:
\bea
m_{\psi\pm}^2&\equiv& V_{\psi\pm}= 2 \kappa^2\,{\rm sn}^2(\omega\sigma+\K;k^2)\pm\kappa\omega{\rm cn}(\omega\sigma+\K;k^2){\rm dn}(\omega
\sigma+\K;k^2)
\ea 

Even though the potentials of different modes look very different, with the transformation:
\bea
{\tilde k}=\frac{2 \sqrt{k}}{1+k}
\label{landen}
\ea
all four potentials can be put into the "canonical" ${\rm sn^2}$ form:
\bea
\op_{\beta}&=&\partial_\tau^2-\dd+\frac{4 \K^2}{\pi^2}\,2 k^2\,{\rm sn}^2(\frac{2 \K}{\pi}\sigma+\K;k^2)\nn
\op_{\phi}&=&\partial_\tau^2-\dd+ \frac{4 \tK^2}{\pi^2}\,2\tk^2\,{\rm sn}^2(\frac{2 \tK}{\pi}\sigma+i\tK^{\prime};\tk^2)-\frac{4 \tK^2 \tk^2}{\pi^2}\nn
\op_{\psi\pm}&=&\partial_\tau^2-\dd+\frac{ \tK^2}{\pi^2}\,2\tk^2\,{\rm sn}^2( \frac{\tK}{\pi}\sigma+\tK\mp\frac{\tK}{2};\tk^2)-\frac{\tK^2 \tk^2}{\pi^2}\nn
\op_{\theta}&=&\partial_\tau^2-\dd
\label{operators}
\ea
This transformation (\ref{landen}) is known as the {\it Landen transformation} and we refer the reader to \cite{abram} and also to the appendices in \cite{Beccaria:2010ry} for the details of it. This canonical form is recognized as single-gap Lam\'e form \cite{whittaker,algebro}, which has the implication that the fluctuation problem can be solved analytically, and the corresponding fluctuation determinant can be computed in closed-form \cite{Beccaria:2010ry}.

The space dependent $\beta$ and the fermion masses ($m_\beta$ and $m_\psi$) are basically periodic arrays of kink anti-kinks. In the short string limit, 
$k\rightarrow0$ they become sinusoidal with vanishing amplitude in agreement with the flat-space limit where the fluctuations are massless. Whereas in 
the long string limit, $k\rightarrow1$, the period goes to infinity and the masses are approximately homogeneous with a $k$ dependence $m_
\beta^2\approx2\kappa^2\approx2\omega^2,m_\psi^2\approx\kappa^2\approx\omega^2$.  On the other hand, the $\phi$ mass is always singular at the 
turning points of the string ($\sigma=\pi/2,3\pi/2$) as mentioned. Nevertheless in the long string limit the $\phi$ mass is also almost homogeneous with 
$m_\phi^2\approx4\kappa^2$ and the singularities are isolated in the turning points.   

The final step before solving the fluctuation problem is to make use of the fact that the fluctuations are static and reduce the partial differential operators to one dimensional ordinary differential 
operators by Fourier transforming the $\tau$ coordinate ($\partial^2_\tau \rightarrow-\partial^2_\tau \rightarrow \Omega^2$). 

The two linearly independent solutions of the single-gap Lam\'e equation;
\bea
[-\dd+2k^2\,m^2\,{\rm sn^2}(m\sigma;k^2)]f(m\sigma)=\Ld f(m\sigma)
\label{Lame}
\ea
are given by \cite{whittaker}:
\bea
f_\pm(m\sigma)=\frac{H(m\sigma\pm\alpha)}{\Theta(m\sigma)}e^{\mp m \sigma Z(\alpha)}
\label{lamesolution}
\ea
with H, $\Theta$ and Z being Jacobi eta, theta and zeta functions. They are quasi periodic under $m\sigma\rightarrow m\sigma+2 \K$:
\bea
f_\pm(m\sigma+2\K)=-f_\pm(m\sigma)e^{\mp 2\K Z(\alpha)}
\label{quasi}
\ea
The parameter $\alpha$ is called the spectral parameter and takes values on the \textit{fundamental rectangle} defined by the region in complex planed 
bordered by the points $0$,$\K$,$i\K^\prime$ and $\K+i\K^\prime$. It also implicitly characterizes the eigenvalue in (\ref{Lame}) through the relation:
\bea
\Ld(\alpha)=m^2(k^2+{\rm dn^2}(\alpha;k^2))
\label{spectral_energy}
\ea

By shifting the wavefunction by one period (say $L$), one defines the Bloch momentum $p$:
\bea
f(x+L)=f(x)e^{ipL}
\ea
In the Lam\'e case, the Bloch momentum can also be parametrized by the spectral parameter $\alpha$, and from (\ref{lamesolution}) we find
\begin{eqnarray}
p(\alpha)=\frac{2 i \K Z(\alpha)}{L}+\frac{\pi}{L}
\end{eqnarray}
 As a result, together with the relation 
(\ref{spectral_energy}), one can write the dispersion relation implicitly, yielding the Bloch momentum as a function of the energy eigenvalue: $p=p(\Lambda)$. In the remaining part of this section we will first investigate the properties of the 
spectrum in the spectral language and then obtain the same results by analyzing the dispersion relation in the large energy limit.

\subsection{Spectral representation of the eigenvalues}\label{spectral}
 
In general, imposing periodic boundary conditions with period $2\pi$, implies the quantization\footnote{The extra minus sign is just for future convenience 
in the notation and has no physical signiÞcance.} of the Bloch momentum $p$: 
\bea
p =-n \quad\quad,\quad  n\in \mathbb{N}
\label{mom_quant}
\ea
If we think of the Bloch momentum as a function of $\alpha$, (\ref{mom_quant}) reads as a quantization condition for $\alpha$. Our main strategy in this 
subsection is to develop a semiclassical approximation scheme to find discrete the set of $\alpha$ values such that (\ref{mom_quant}) holds and then 
evaluate (\ref{spectral_energy}) on these values to construct the spectrum.

\subsubsection{Bosonic modes}\label{spec_bos}

\begin{center} (i) \textit{$\beta$ mode:} \end{center}

The amplitude $m$ of the bosonic potentials is $\frac{2 \K}{\pi}$, where the elliptic quarter-period $\K$ is evaluated at $k^2$ for the $\beta$ modes, and at the Landen transformed (\ref{landen}) value $\tk^2$ for the $\phi$ mode. Together with the closed string condition $\sigma\sim\sigma+2\pi$, the resulting shift by one period is $mL=4 \K$. The Bloch momentum for the 
bosonic modes is then found to be:
\bea
p(\alpha)=i m Z(\alpha)=i\frac{2 \K}{\pi}Z(\alpha)
\label{palpha}
\ea
Quantization of $\alpha$ can be achieved by solving the equation;
\bea
Z(\alpha_n)=\frac{i n \pi}{2 \K}
\label{spectral_zeroes}
\ea
In general, the inverse of the zeta function cannot be expressed in an analytical way except for the following special cases:
\bea
\alpha_I=\K+i\K^\prime \quad\quad \alpha_{II}=\K\quad\quad \alpha_{III}=0 \nn
Z(\alpha_I)=\frac{i\pi}{2 \K}\quad\quad Z(\alpha_{II})=Z(\alpha_{III})=0
\ea

\begin{figure}[h]
\center
\includegraphics[scale=0.65]{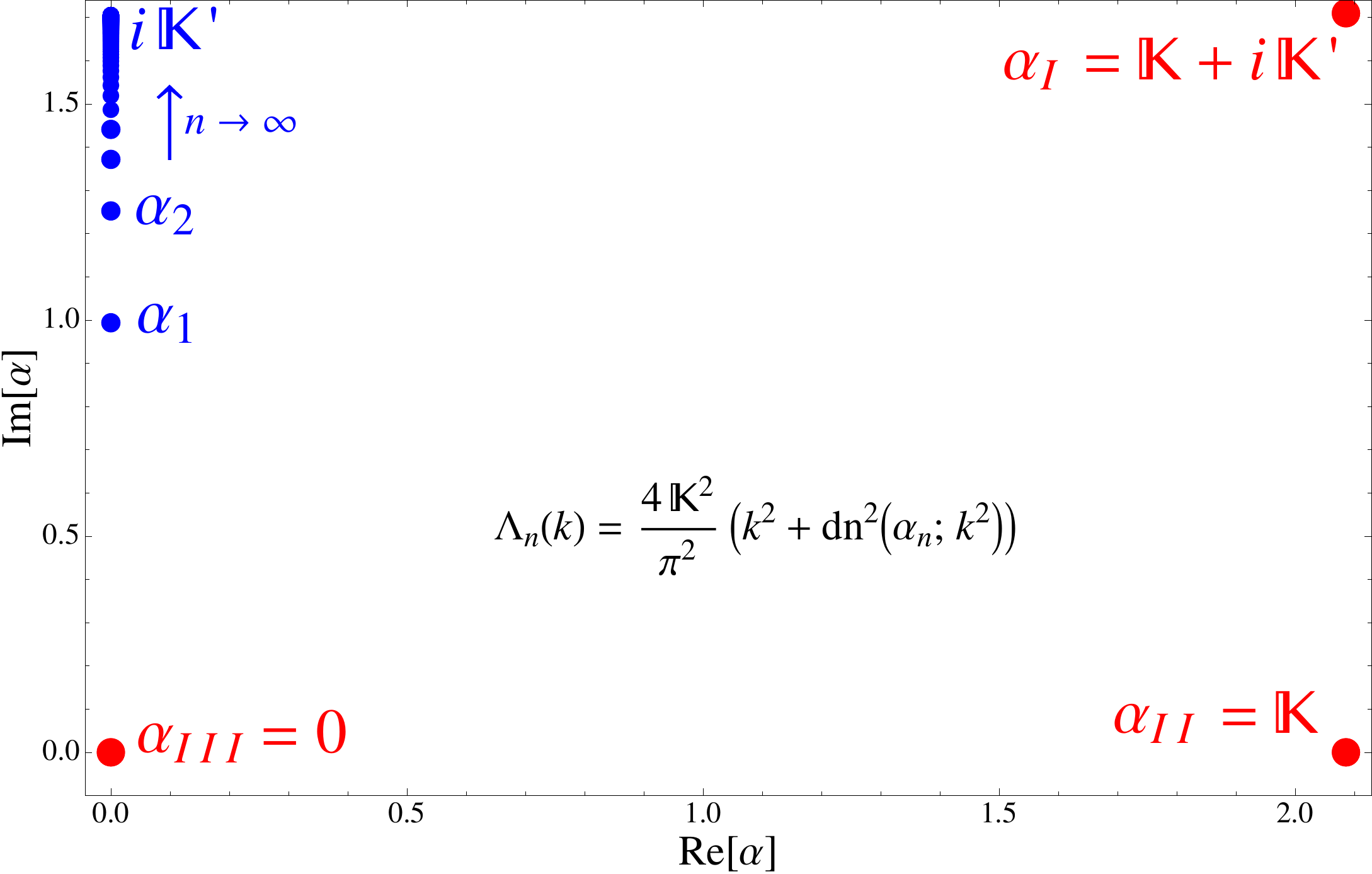}
\caption{The solutions of $Z(\alpha_n)=\frac{i n \pi}{2 \K}$ for $k$=0.5 on the fundamental rectangle and the fluctuation eigenvalue as a function of the 
spectral parameter $\alpha$. The three red dots ($\alpha_{I,II,III}$) correspond to analytically known band edges and the blue dots ($\alpha_n$) 
correspond to quantized values of the continuum energies. Observe that they accumulate around $i \K^\prime$ which is the high energy regime.}
\label{spectral_zeroes_fig}
\end{figure}
These points precisely correspond to three of the points defining the fundamental rectangle. Plugged into (\ref{spectral_energy}), they also characterize 
the band edges, which are the lowest three eigenvalues:
\bea
\Ld_{I}=\frac{4\K^2}{\pi^2}k^2 \quad\quad \Ld_{II}=\frac{4\K^2 }{\pi^2}\quad\quad \Ld_{III}=\frac{4\K^2}{\pi^2}(1+k^2)
\label{beta_edges}
\ea 

The remaining solutions of (\ref{spectral_zeroes}) are pure imaginary and lie in the interval $[0,i \K^\prime)$. Furthermore both $i Z$ and ${\rm dn^2}$ are 
monotonously increasing functions of $\alpha$ in this interval and $ Z(\alpha),{\rm dn^2}\rightarrow\infty$ as $\alpha\rightarrow i\K^\prime$. Thus the 
spectral values describing the highly excited (large $n$) energy states are accumulated around the limit point $i\K^\prime$. We now expand the zeta 
function around this point. Physically this is just the WKB expansion evaluated on the spectral plane. Defining the expansion parameter $\gamma_n$ as;
\bea
\alpha_n=i \K^\prime-i \gamma_n \nonumber 
\ea
and using the well known relation between the elliptic integrals we write $Z$ as an integral of ${\rm dn^2}$:
\bea
Z(\gamma_n)&=&{E}(\gamma_n)-\frac{\E}\K(i \K^\prime-i \gamma_n)=\int_0^{i\K^\prime-i\gamma_n}{\mbox {dn}}^2 (x)dx-\EK(i \K^\prime-i \gamma_n) \nn
&=&i\left(-\int_{\K^\prime}^{\gamma_n}{\mbox {dn}}^2 (i \K^\prime-i x)dx+\EK(\gamma_n-\K^\prime)\right)
\label{zeta}
\ea
${\rm dn^2}$ has the following Laurent expansion around $i\K^\prime$:
\bea
{\mbox {dn}}^2 (i \K^\prime-i x)=\frac{1}{x^2}+\sum_{j=0}^\infty \, c_{2j}(k)x^{2j}
\label{dn_expansion}
\ea
where $c_{2j}(k)$ s are polynomials in $k^2$. The first few terms in the expansion are:
\bea
c_0&=&\frac{2 - k^2}{3} \nn
c_2&=&\frac{1}{15} (1 - k^2 + k^4)  \nn
c_4&=& \frac{1}{189} (-2 + 3 k^2+ 3 k^4 - 2 k^6)\nn
c_6&=& \frac{1}{675} (1 - 2 k^2+ 3 k^4 - 2 k^6 + k^8)
\ea
After substituting the expansion (\ref{dn_expansion}) into (\ref{zeta}) and integrating, we obtain the desired expansion\footnote{The constant term $
\frac{\pi}{2 \K}$ comes from the antiderivative of ${\rm dn^2}$ evaluated at $\K^{\prime}$}:
\bea
Z(\gamma_n)=i\left[\frac{1}{\gamma_n}+\left(\frac\E \K-c_0\right)\gamma_n-\sum_{j=1}^{\infty}\frac{c_{2j}}{2j+1}\gamma_n^{2j+1}-\frac{\pi}{2\K} \right]
\label{zeta_exp}
\ea
Plugging (\ref{zeta_exp}) into (\ref{spectral_zeroes}) and truncating the sum at an arbitrary $j$=N yields a polynomial of order 2N+2 in $\beta_n$, which is in principle 
numerically soluble. Nevertheless, in order to get an analytical result let us make the following ansatz for $\gamma_n$:
\bea
\gamma_n(k)=\frac{2\K}{\pi (n+1)}+\sum_{j=1}^{\infty}\frac{b_{2j+1}(k)}{(n+1)^{2j+1}} \quad\quad n\in \mathbb Z^+
\label{ansatz}
\ea 
and insert into (\ref{zeta_exp}) to get:
\bea
Z(n)=\frac{i n \pi}{2 \K} + \sum_{j=0}^{\infty} \frac{P_{2j+3}}{(n+1)^{2j+1}} 
\ea
where $P$s are known polynomials in $b_k$s. For any $j$, $P_{2j+3}$ is linear in $b_{2j+3}$ and contains only powers of $b_m$s with $m<2j+3$. As a 
result it is possible to solve $P_{2j+3}=0$ up to any desired order inductively. The first few $b_j$s are:
\bea
b_3&=&\frac{1}{3}\left(\frac{2\K}{\pi}\right)^3\left(3\EK-2+k^2\right) \nn
b_5&=&\frac{1}{15}\left(\frac{2\K}{\pi}\right)^5\left( 30\left(\EK\right)^2+20(-2+k^2)\EK+(13+k^2(-13+3k^2))\right) \nn
b_7&=&\frac{1}{105}\left(\frac{2\K}{\pi}\right)^7\left(525\left(\EK\right)^3+525(-2+k^2 )\left(\EK\right)^2 \right.\nn
&&\left. +7 (98+k^2 (-98+23k^2 ))\EK+(-2+k^2 )(73+k^2(-73+15 k^2)) \right)
\ea
The pattern is simple. Each $b_{2j+1}$ is a polynomial of $\EK$ and $k^2$ of order $j$ with an overall factor of $\left(\frac{2 \K}{\pi}\right)^{2j+1}$ which is recognized as the amplitude of the potential. It should 
be kept in mind that the overall $\K$ factor diverges as $k\rightarrow1$ and the approximation breaks down for $k$ extremely close to 1. We will come 
back to this point after we discuss the eigenvalues. 
\begin{table}[H]
\center
\begin{tabular}{lclllll}
\cline{3-7}
& & \multicolumn{5}{|c|}{$n$} \\ \cline{1-7}
 \multicolumn{1}{|c|}{$k$} & \multicolumn{1}{|c|}{to order:}& \multicolumn{1}{|c|}{$1$}& \multicolumn{1}{|c|}{$2$}& \multicolumn{1}{|c|}{$3$}& 
\multicolumn{1}{|c|}{$4$}& \multicolumn{1}{|c|}{$5$}\\ \cline{1-7}
\multicolumn{1}{|c|}{\multirow{6}{*}{$0.1$}} &
\multicolumn{1}{|c|}{exact} &\,4.40477\,\,& 9.40102\,\, & 16.400\,\, & 16.4000\,\, &  \multicolumn{1}{l|}{25.3995}    \\ \cline{2-7}
\multicolumn{1}{|c|}{}                        &
\multicolumn{1}{|c|}{$(n+1)^{-2}$} &\,4.40328\,\,& 9.40077\,\, & 16.399\,\, & 16.3999\,\, &  \multicolumn{1}{l|}{25.3995}    \\ \cline{2-7}
\multicolumn{1}{|c|}{}                        &
\multicolumn{1}{|c|}{$(n+1)^{-4}$}  &\,4.40440\,\,& 9.40099\,\, & 16.400\,\, & 16.4000\,\, &  \multicolumn{1}{l|}{25.3995}    \\ \cline{2-7}
\multicolumn{1}{|c|}{}                        &
\multicolumn{1}{|c|}{$(n+1)^{-6}$}  &\,4.40468\,\,& 9.40101\,\, & 16.400\,\, & 16.4000\,\, &  \multicolumn{1}{l|}{25.3995}    \\ \cline{2-7}
\multicolumn{1}{|c|}{}                        &
\multicolumn{1}{|c|}{$(n+1)^{-8}$} &\,4.40475\,\,& 9.40102\,\, & 16.400\,\, & 16.4000\,\, &  \multicolumn{1}{l|}{25.3995}    \\ \cline{1-7}
\multicolumn{1}{|c}{}  & & & & & &  \multicolumn{1}{l|}{$ $} \\ \cline{1-7}
 \multicolumn{1}{|c|}{$k$} & \multicolumn{1}{|c|}{to order:}& \multicolumn{1}{|c|}{$1$}& \multicolumn{1}{|c|}{$2$}& \multicolumn{1}{|c|}{$3$}& 
\multicolumn{1}{|c|}{$4$}& \multicolumn{1}{|c|}{$5$}\\ \cline{1-7}
\multicolumn{1}{|c|}{\multirow{6}{*}{$0.5$}} &
\multicolumn{1}{|c|}{exact} &\,5.49670\,\,& 10.4585\,\, & 17.4475\,\, & 26.4427\,\, &  \multicolumn{1}{l|}{37.4402}    \\ \cline{2-7}
\multicolumn{1}{|c|}{}                        &
\multicolumn{1}{|c|}{$(n+1)^{-2}$} &\,5.48278\,\,& 10.4561\,\, & 17.4468\,\, & 26.4424\,\, &  \multicolumn{1}{l|}{37.4401}    \\ \cline{2-7}
\multicolumn{1}{|c|}{}                        &
\multicolumn{1}{|c|}{$(n+1)^{-4}$}  &\,5.49394\,\,& 10.4583\,\, & 17.4475\,\, & 26.4427\,\, &  \multicolumn{1}{l|}{37.4402}    \\ \cline{2-7}
\multicolumn{1}{|c|}{}                        &
\multicolumn{1}{|c|}{$(n+1)^{-6}$}  &\,5.49624\,\,& 10.4585\,\, & 17.4475\,\, & 26.4427\,\, &  \multicolumn{1}{l|}{37.4402}    \\ \cline{2-7}
\multicolumn{1}{|c|}{}                        &
\multicolumn{1}{|c|}{$(n+1)^{-8}$} &\,5.49665\,\,& 10.4585\,\, & 17.4475\,\, & 26.4427\,\, &  \multicolumn{1}{l|}{37.4402}    \\ \cline{1-7}
\multicolumn{1}{|c}{}  & & & & & &  \multicolumn{1}{l|}{$ $} \\ \cline{1-7} \multicolumn{1}{|c|}{$k$} & \multicolumn{1}{|c|}{to order:}& \multicolumn{1}{|c|}{$1$}& 
\multicolumn{1}{|c|}{$2$}& \multicolumn{1}{|c|}{$3$}& \multicolumn{1}{|c|}{$4$}& \multicolumn{1}{|c|}{$5$}\\ \cline{1-7}
\multicolumn{1}{|c|}{\multirow{6}{*}{$0.9$}} &
\multicolumn{1}{|c|}{exact} &\,8.64532\,\,& 13.4485\,\, & 20.3838\,\, & 29.3549\,\, &  \multicolumn{1}{l|}{40.3394}    \\ \cline{2-7}
\multicolumn{1}{|c|}{}                        &
\multicolumn{1}{|c|}{$(n+1)^{-2}$} &\,8.60953\,\,& 13.4404\,\, & 20.3812\,\, & 29.3538\,\, &  \multicolumn{1}{l|}{40.3389}    \\ \cline{2-7}
\multicolumn{1}{|c|}{}                        &
\multicolumn{1}{|c|}{$(n+1)^{-4}$}  &\,8.65301\,\,& 13.4490\,\, & 20.3839\,\, & 29.3549\,\, &  \multicolumn{1}{l|}{40.3394}    \\ \cline{2-7}
\multicolumn{1}{|c|}{}                        &
\multicolumn{1}{|c|}{$(n+1)^{-6}$}  &\,8.64873\,\,& 13.4486\,\, & 20.3838\,\, & 29.3549\,\, &  \multicolumn{1}{l|}{40.3394}    \\ \cline{2-7}
\multicolumn{1}{|c|}{}                        &
\multicolumn{1}{|c|}{$(n+1)^{-8}$} &\,8.64553\,\,& 13.4485\,\, & 20.3838\,\, & 29.3549\,\, &  \multicolumn{1}{l|}{40.3394}    \\ \cline{1-7}
\multicolumn{1}{|c}{}  & & & & & &  \multicolumn{1}{l|}{$ $} \\ \cline{1-7} \multicolumn{1}{|c|}{$k$} & \multicolumn{1}{|c|}{to order:}& \multicolumn{1}{|c|}{$1$}& 
\multicolumn{1}{|c|}{$2$}& \multicolumn{1}{|c|}{$3$}& \multicolumn{1}{|c|}{$4$}& \multicolumn{1}{|c|}{$5$}\\ \cline{1-7}
\multicolumn{1}{|c|}{\multirow{6}{*}{$0.99$}} &
\multicolumn{1}{|c|}{exact} &\,14.9029\,\,& 19.3857\,\, & 26.1877\,\, & 35.0916\,\, &  \multicolumn{1}{l|}{46.0383}    \\ \cline{2-7}
\multicolumn{1}{|c|}{}                        &
\multicolumn{1}{|c|}{$(n+1)^{-2}$} &\,15.0425\,\,& 19.4159\,\, & 26.1966\,\, & 35.0951\,\, &  \multicolumn{1}{l|}{46.0400}    \\ \cline{2-7}
\multicolumn{1}{|c|}{}                        &
\multicolumn{1}{|c|}{$(n+1)^{-4}$}  &\,14.9224\,\,& 19.3922\,\, & 26.1891\,\, & 35.0920\,\, &  \multicolumn{1}{l|}{46.0385}    \\ \cline{2-7}
\multicolumn{1}{|c|}{}                        &
\multicolumn{1}{|c|}{$(n+1)^{-6}$}  &\,14.8178\,\,& 19.3830\,\, & 26.1875\,\, & 35.0916\,\, &  \multicolumn{1}{l|}{46.0383}    \\ \cline{2-7}
\multicolumn{1}{|c|}{}                        &
\multicolumn{1}{|c|}{$(n+1)^{-8}$} &\,14.8726\,\,& 19.3851\,\, & 26.1877\,\, & 35.0916\,\, &  \multicolumn{1}{l|}{46.0383}    \\ \cline{1-7}
\end{tabular}
\caption{Comparison of the fluctuation eigenvalues $\Lambda_n$ of the $\beta$ mode calculated up to various orders in $(n+1)^{-2}$ and for various 
values of the elliptic parameter $k$ to the numerically evaluated exact values. Observe the persistence of the remarkable agreement even for lowest 
eigenvalue $n=1$, especially around small $k$. The approximation breaks down very slowly as k approaches 1. The results for other modes ($\phi$, $\psi
$) are similar.}
\label{table}
\end{table}

We now basically have all the information we need to construct the eigenvalues in the WKB limit where we write $\Ld_n$ as a series in $(n+1)^{-2}$. $
\Ld(\alpha)$ is basically ${\rm dn}^2(\alpha)$, so it suffices to use (\ref{ansatz}) in (\ref{dn_expansion}) to obtain the WKB series of the $\beta$ mode. 
\bea
\Ld_{\beta\,n}&=&\frac{4 \K^2}{\pi^2}(k^2+{\rm dn}^2(\alpha_n;k^2))\nn&&=(n+1)^2+2d_0(k)+\sum_{j=1}^{\infty}\frac{d_{2j}(k)}{(n+1)^{2j}} \quad\quad n\in \mathbb 
Z^+
\label{beta_energy}
\ea  
with the coefficients; 
\bea
d_0(k)&=&\left(\frac{2\K}{\pi}\right)^2\left(-\EK+1\right) \nn
d_2(k)&=&\frac{1}{3}\left(\frac{2\K}{\pi}\right)^4\left(-3\left(\EK\right)^2-2(-2+k^2)\EK+(-1+k^2)\right) \nn
d_4(k)&=&\frac{1}{5}\left(\frac{2\K}{\pi}\right)^6\left(-10\left(\EK\right)^3-10(-2+k^2 )\left(\EK\right)^2-2 (6+k^2 (-6+k^2 ))\EK\right. \nn
&&\left.+(-2+k^2 )(-1+k^2) \right)\nn
d_6(k)&=&\frac{1}{63}\left(\frac{2\K}{\pi}\right)^8\left(-315\left(\EK\right)^4-420(-2+k^2 )\left(\EK\right)^3-7 (113+k^2 (-113+23k^2 ))\left(\EK\right)^2 \right. 
\nn &&+ \left.2 (-2 +k^2) (76 + k^2 (-76 + 9 k^2))\EK+(-1 +k^2) (38 + k^2 (-38 + 9 k^2) \right)
\label{dns}
\ea

\begin{center} (ii) \textit{$\phi$ mode:} \end{center}

By looking at the fluctuation operator of $\phi$, $\op_\phi$ in (\ref{operators}) one realizes that, it can be obtained by Landen transforming ($k\rightarrow\tk$) 
and shifting the $\beta$ operator by $-\frac{4 \tK \tk^2}{\pi^2}$. Thus the eigenvalues of $\beta$ and $\phi$ fluctuations are connected in the following way: 
\bea
\Ld_\phi(k)=\Ld_\beta(\tk)-\frac{4 \tK \tk^2}{\pi^2}
\label{phi_beta}
\ea
In particular, the band edges are found to be:
\bea
\Ld_{\phi,\,I}&=&0 \quad\quad \Ld_{\phi,\,II}=\frac{4\tK^2 }{\pi^2}(1-\tk^2)\quad\quad \Ld_{\phi,\,III}=\frac{4\tK^2}{\pi^2}\nn
\Ld_{\phi,\,I}&=&0 \quad\quad \Ld_{\phi,\,II}=\frac{4 \K^2 }{\pi^2}(1-k)^2\quad\quad \Ld_{\phi,\,III}=\frac{4\K^2}{\pi^2}(1+k)^2
\label{phi_edges}
\ea  
where in the second line we made an inverse Landen transformation and wrote the band edges as functions of $k$. Next, applying the identity (\ref{phi_beta}) to the series $\Lambda_{\beta\,n}$ (\ref{beta_energy}) yields the WKB series for $\Ld_{\phi,\,n}$: 
\bea
\Ld_{\phi,\,n}&=&\frac{4 \tK^2}{\pi^2}{\rm dn}^2(\alpha_n;\tk^2)\nn&&=(n+1)^2+2d_0(\tk)-\frac{4 \tK \tk^2}{\pi^2}+\sum_{j=1}^{\infty}\frac{d_{2j}(\tk)}{(n+1)^{2j}}\quad
\quad n\in \mathbb Z^+
\label{phi_energy0}
\ea  
Again, an inverse Landen transformation of the constant term simplifies the expression further;
\bea
2d_0(\tk)-\frac{4 \tK \tk^2}{\pi^2}=4d_0(k)
\ea 
It should be stressed that this is precisely the same function as in (\ref{dns}).
We also give the first few terms in (\ref{phi_energy0}) as functions of $k$ for completeness;
\bea
d_2(\tk)&\equiv&\tilde{d_2}(k)=\frac{2}{3}\left(\frac{2\K}{\pi}\right)^4\left(-3\left(\EK\right)^2-(-5+k^2)\EK+2(-1+k^2)\right) \nn \nn
d_4(\tk)&\equiv&\tilde{d_4}(k)=\frac{2}{5}\left(\frac{2\K}{\pi}\right)^6\left(-20\left(\EK\right)^3-10(-5+k^2 )\left(\EK\right)^2-(41 +k^2 (-26 + k^2))\EK \right.\nn
&& \left. +(-1+k^2 )(-11+3k^2) \right)\nn\nn
d_6(\tk)&\equiv&\tilde{d_6}(k)=\frac{2}{63}\left(\frac{2\K}{\pi}\right)^8\left(-1260\left(\EK\right)^4-840(-5+k^2 )\left(\EK\right)^3\right.\nn
&& \left. -7 (743+k^2 (-389+23k^2 ))
\left(\EK\right)^2 +12 (2833+k^2) (-2811 + k^2 (563- 9 k^2))\EK\right. \nn
&&\left.+4(-1 +k^2) (143 + k^2 (-80 + 9 k^2) \right)
\ea
where all the elliptic functions in the right-hand side are evaluated at $k$. The final form of $\Ld_{\phi,\,n}$ is then given as:
\bea
\Ld_{\phi,\,n}=(n+1)^2+4d_0(k)+\sum_{j=1}^{\infty}\frac{\td_{2j}(k)}{(n+1)^{2j}}\quad\quad n\in \mathbb Z^+
\label{phi_energy}
\ea

An important observation is that both in (\ref{beta_energy}) and (\ref{phi_energy}), all the coefficients $d_j$ vanish  at $k$=0 so $\Ld_n \sim (n+1)^2$. This is 
the reflection of the flat-space limit, where all the fluctuations are massless, in the WKB scheme and the $\ord(n^{-2m})$ corrections can be thought as 
curvature corrections.

Another point worth mentioning is the applicability of this WKB approximation. As mentioned, the approximation breaks down in the large spin limit ($k
\rightarrow1$) and the coefficients $d_j$ diverge due to the overall factors of powers of $\K$. In general the divergence is logarithmic and the $(n+1)^{-2}$ 
expansion is significantly accurate (see table \ref{table}). Around the $k=1$ limit, the series behave as an asymptotic series and the 
accuracy depends on the number of terms taken. When we compute the one loop energy correction, we sum over $n$ and the asymptotic behavior of the 
series requires careful treatment.
With these limitations, this WKB scheme, which works very well in the short-medium spin regime, can 
be thought as a complementary approximation to the well known long spin one where the potentials are taken as homogeneous and the eigenvalues are 
given by $\Ld_\beta \approx n^2+4\kappa^2$ etc.. \cite{Frolov:2002av} 

Finally, the WKB approximation is expected to work well in the large eigenvalue regime. It is remarkable that our results are extremely accurate even for the 
low energy eigenvalues (see table \ref{table}). Again, the agreement gets better especially in the small $k$ region but extends to medium even large $k$.

\subsubsection{Fermionic modes}
From (\ref{operators}), we see that the amplitude of the fermionic potential is $\frac{\tK}{\pi}$ and with $L=2\pi$, the shift under one period is $2 \tK$. The 
overall minus sign in (\ref{quasi}) brings an extra $\frac{1}{2}$ term and sets the fermionic Bloch momentum as:
\bea
p(\alpha)=i\frac{\tK}{\pi}Z(\alpha)+\frac{1}{2}
\ea
After quantization (\ref{mom_quant}) we obtain the following equation for the spectral parameter;
\bea
Z(\alpha_n)=\frac{i \pi}{\tK}(n+\frac{1}{2})
\label{z_psi}
\ea
Comparing (\ref{z_psi}) with (\ref{spectral_zeroes}) one can relate the spectral parameter of the $\psi$ mode with $\phi$ mode as:
\bea
\alpha_{\psi,\,n}=\alpha_{\phi,\,2n+1}
\label{phipsi}
\ea
or as:
\bea
\gamma_{\psi,\,n}(\tk)=\frac{\K}{\pi(n+1)}+\sum_{j=1}^{\infty}\frac{b_{2j+1}(\tk)}{2^{2j+1}(n+1)^{2j+1}} \quad\quad n\in \mathbb N
\ea 
which is a reflection of the oscillation theorem \cite{magnus}. There is also one analytically known solution of (\ref{z_psi}) which is $\alpha_{\psi,I}=\tK+i\tK^\prime$ and the corresponding state is the fermionic zero 
mode. It is also straightforward to calculate the higher eigenvalues by using (\ref{dn_expansion}) and (\ref{phipsi});
\bea
\Ld_{\psi,\,n}&=&\frac{\tK^2}{\pi^2}{\rm dn}^2(\alpha_{\psi,\,n};\tk^2)=(n+1)^2+\frac{d_0(\tk)}{2}-\frac{\tK \tk^2}{\pi^2}+\sum_{j=1}^{\infty}\frac{d_{2j}(\tk)}{2^{2j
+2}(n+1)^{2j}}\quad\quad n\in{\mathbb N}\nn
&=&(n+1)^2+d_0(k)+\sum_{j=1}^{\infty}\frac{\td_{2j}(k)}{2^{2j+2}(n+1)^{2j}}\quad\quad n\in{\mathbb N}\nn
\label{psi_energy}
\ea

Before concluding the section, we note that the highly excited states (very large $n$) behave like massless excitations  $\Ld\sim n^2$ as a typical behavior of the WKB construction. 
When we sum over the eigenvalues to calculate the one loop energy, the 8 bosonic degrees (2 $\beta$ + 1 $\phi$ + 5 $\theta$ ) and the 8 fermionic 
degrees will cancel each other out, resulting in a UV convergent expression, as a result of the underlying supersymmetry of the theory.

\subsection{Large energy expansion}
\label{heat_kernel}
In the previous section we used the spectral representation of the fluctuation eigenvalues to construct the WKB series. This is possible because we have
an explicit closed- form expression for the quasi-momentum of each mode. We now use another method to derive the same result. The advantage of this method is that it is more general, in the sense that it can be 
applied to more general fluctuation problems whose exact solutions are not known. The idea is to use the dispersion relation $dp=\frac{dp(\Ld)}{d\Ld}\,d\Ld$ and impose the momentum 
quantization (\ref{mom_quant}) to solve\footnote{This procedure can also be thought as a special case of imposing Bohr-Sommerfeld type quantization performed on an hyperelliptic curve 
which characterizes "spiky" strings with arbitrary number of folds \cite{dorey}. Here our string has 2 folds and can be described by an elliptic curve.} for $\Ld_n$. In general the momentum $p$ and energy $\Ld$ are functions of $\alpha$;
\bea
p_\beta(\alpha)&=&\frac{2 \K}{\pi}iZ(\alpha;k) \,\,\,\qquad\quad\Ld_\beta(\alpha)=\frac{4 \K^2}{\pi^2}(k^2+{\rm dn}^2(\alpha;k^2))\nn
p_\phi(\alpha)&=&\frac{2 \tK}{\pi}iZ(\alpha;\tk) \,\,\,\qquad\quad\Ld_\phi(\alpha)=\frac{4 \tK^2}{\pi^2}{\rm dn}^2(\alpha;\tk^2)\nn
p_\psi(\alpha)&=&\frac{\tK}{\pi}iZ(\alpha;\tk)+\frac{1}{2} \qquad\Ld_\psi(\alpha)=\frac{\tK^2}{\pi^2}{\rm dn}^2(\alpha;\tk^2)
\label{dispersion}
\ea
Making use of the properties of the elliptic functions we find (for the $\beta$ mode);
\bea
\frac{dp}{d\Ld}=\frac{dp}{d\alpha}\frac{d\alpha}{d\Ld}&=&\frac{\pi}{2\K}\frac{{\rm dn}^2(\alpha)-\EK}{2\,{\rm dn}(\alpha) \sqrt{{\rm dn}^2(\alpha)-1}\sqrt{{\rm 
dn}^2(\alpha)-1+k^2}}\nn
&=&\frac{\Ld-\frac{4 \K^2}{\pi^2}\left(k^2+\EK\right)}{2\sqrt{(\Ld-\Ld_I)(\Ld-\Ld_{II})(\Ld-\Ld_{III})}}
\ea 
where $\Ld_I,\,\Ld_{II},\,\Ld_{III}$ are the band edges defined in (\ref{beta_edges}). It is also useful to observe that the constant piece in the numerator can 
be written as;
\bea
\mu_\beta\equiv\frac{4 \K^2}{\pi^2}\left(k^2+\EK\right)=\frac{1}{2}\left(\Ld_I+\Ld_{II}+\Ld_{III}-\frac{1}{2\pi}\int_0^{2\pi}V_\beta(\sigma)d\sigma\right)
\ea
Similarly, the $\phi$ and $\psi$ modes have the same form of the dispersion relation;
\bea
\frac{dp_f}{d\Ld}=\frac{\Ld-\mu_f}{2\sqrt{(\Ld-\Ld_{f,\,I})(\Ld-\Ld_{f,\,II})(\Ld-\Ld_{f,\,III})}}
\label{dispersion2}
\ea 
where $f$ denotes different modes. $\mu_f$s for $\phi$ and $\psi$ modes are given by:
\bea
\mu_\phi=4\mu_\psi=\frac{4 \tK^2}{\pi^2}\frac{\tE}{\tK}
\ea
We also define:
\bea
\Ld_{\psi,\,I}=0 \qquad \Ld_{\psi,\,II}=\frac{1}{4}\Ld_{\phi,\,II} \qquad \Ld_{\psi,\,III}=\frac{1}{4}\Ld_{\phi,\,III}
\ea
But it should be kept in mind that $\Ld_{\psi,\,II}$ and $\Ld_{\psi,\,III}$ are {\textit not} eigenvalues of the ferimonic fluctuation operators. They are just $
\Ld(\alpha)$s evaluated at the special values $\alpha_{II}$ and $\alpha_{III}$. The only analytically known fermionic eigenvalue is zero. 
At this point we reformulate the WKB expansion by expanding (\ref{dispersion2}) around large $\Ld$. The expansion makes it possible to integrate the 
right-hand side with respect to $\Ld$ and obtain a series that makes the dispersion relation $p=p(\Ld)$ explicit. Then we impose the momentum 
quantization to work out the eigenvalues.

Our main expansion is:
\bea
\frac{\Ld-\mu}{2\,\sqrt{(\Ld-\Ld_{I})(\Ld-\Ld_{II})(\Ld-\Ld_{III})}}\approx\frac{1}{2 \sqrt{\Ld }}+\frac{\Ld_I+\Ld_{II}+\Ld_{III}-2 \mu }{4\, \Ld ^{3/2}}+\ord(\Ld^{-5/2})
\label{largeE}
\ea
and after integrating with respect to $\Ld$, we write and quantize the momentum as; 
\bea
p(\Ld)&\approx&\sqrt{\Ld}-\frac{\Ld_I+\Ld_{II}+\Ld_{III}-2 \mu }{2\, \Ld ^{1/2}}+\ord(\Ld^{-3/2})-1\nn
&\equiv&n\qquad n\in\mathbb N
\label{mom_largeE}
\ea
where the extra term 1 is an integration constant fixed by the periodic boundary conditions. It is basically the same as the constant term in (\ref{zeta_exp}). 
Also this form of $p(\Ld)$ is basically the dispersion relation of the free fluctuations plus $\Ld^{-j/2}$ corrections as one would expect. We now take the 
form (\ref{beta_energy}) as an \textit{ansatz};
\bea
\Ld_n\equiv(n+1)^2+2d_0+\sum_{j=1}^{\infty}\frac{d_{2j}}{(n+1)^{2j}}
\label{lambda_ans}
\ea
with undetermined coefficients $d_{2j}$. Then we expand $p(\Ld_n)$ given by (\ref{mom_largeE}) up to some particular order in $(n+1)^{-2j}$  and solve $p(\Ld_n)=-n$. The leading 
term is $n+1$. By starting from $j=2$, matching each coefficient of $(n+1)^{-2j}$ with zero and solving for $d_{2j}$, we construct the WKB series 
inductively as we did for $b_{2j}$ in section (\ref{spec_bos}). As a result we find exactly the same $d_{2j}$s given in section (\ref{spec_bos}). The 
advantage of taking such an approach is that it can be applied to arbitrary potentials. In general for an arbitrary potential it is not possible to express the 
dispersion relation even in an implicit way as in (\ref{dispersion}). However one can always make a large $\Ld$ expansion like (\ref{largeE}) and calculate 
the first few series in the WKB expansion easily even though it gets tedious to calculate the higher order coefficients. This procedure can be useful especially for dealing the cases with non-vanishing R-charge where the fluctuations are coupled and not soluble.  

The generic form of (\ref{mom_largeE}) is: 
\bea
p\approx\sqrt{\Ld}-\frac{\langle V \rangle}{2\,\Ld^{1/2}}-\frac{\langle V^2 \rangle}{8\,\Ld^{3/2}}-\frac{\langle 2V^3+V^{\prime\,2}\rangle}{32\,\Ld^{5/2}}+
\ord(\Ld^{-7/2})-1
\label{mom_largeE2}
\ea
Here $\langle V^k \rangle\equiv\frac{1}{L}\int_0^L dx\,(V(x))^k$ and the -1 term is again due to the periodic boundary conditions. This form can be derived from the heat kernel expansion of the corresponding resolvent of the problem. One can check that, evaluated 
on the Lam\'e potentials $V=V_f(\sigma)$,  (\ref{mom_largeE2}) generates the correct coefficients given in (\ref{mom_largeE}). After inverting the equation 
and imposing the quantization on the momentum, we get:
\bea
\Lambda_n = (n+1)^2+\langle V\rangle +\frac{-\langle V^2 \rangle+\langle V \rangle}{4\,(n+1)^2}+\frac{16 \langle V\rangle-6\langle V\rangle\langle 
V^2\rangle+\langle 2V^3+V^{\prime\,2}\rangle}{4\,(n+1)^4}+\ord((n+1)^{-6})
\label{lambda_large_E}
\ea
which is the extended version of the theorem (\ref{magnus}) given in the introduction with some slight modifications. 

First, the leading term $\langle V \rangle$ is present and nonzero in general. It can be realized as the leading order "mass" correction. Note that in the large spin regime using the \textit{amplitude} instead of the mean value of the potential leads to a much better approximation \cite{Frolov:2002av}. The reason for this is, in the large spin regime, the potential $V$ is almost constant with a sharp variation in the end points and it is more accurate to cut off these ends instead of averaging over. This is just a result of the applicability of the derivative expansion as an asymptotic expansion in general \cite{Dunne:1999uy}, and illustrates the nature of the WKB expansion. Each coefficient in (\ref{lambda_large_E}) includes a power of the potential and its derivatives whose order increase as one includes more terms in the series. In order for the derivative expansion to be valid, the potentials should be slowly varying which is violated by the sharply varying edges of the potentials in large spin limit. 

The second point is in the theorem, (\ref{magnus}) the consecutive even and odd eigenvalues behave in the same way, but in our case we do not see this 
behavior. The reason is that, except the band edges, all of the eigenvalues are doubly degenerate (the blue points in fig.\ref{spectral_zeroes_fig}), because the Lam\'e fluctuation potentials are {\it single-gap}, having only one gap in their band spectrum.

Before concluding this section we would like to mention another aspect of the derivative expansion.  One can write down a derivative expansion like 
(\ref{mom_largeE2}) directly for the effective action using the heat kernel expansion.  The variation of the effective action expanded up to $N^{th}$ order, 
with respect to $V$ leads to a differential equation of order $N-2$. In general, these differential equations define an integrable hierarchy known as the KdV hierarchy \cite{algebro}. Each solution of the $N^{th}$ order equation is characterized with a spectrum with $\frac{N}{2}-1$ gaps and solves all the higher order equations  as a particular case\footnote{Pictorially, every general solution to the $m$ gap problem is a particular solution of a $N$ gap problem with $N>m$ and can be obtained by taking $N-m$ band edges to be equal.}. The fluctuation equation (\ref{Lame}) is nothing but the lowest nontrivial equation of the KdV hierarchy. This is the underlying reason why we can sum (\ref{mom_largeE2}) to all orders and express it as (\ref{dispersion}). In other words, each coefficient in (\ref{mom_largeE2}) is governed by a conserved quantity of the hierarchy and can be reduced to form linear in $V$ \cite{Correa:2009xa}. 

Particularly, a similar integrable structure plays an important role in the context of spiky strings \cite{dorey,spiky}. There, the formulation is done through the Lax pair 
and imposing Bohr-Sommerfeld quantization on the hyperelliptic curve defined on a genus $N-2$ Riemann surface\footnote{Compare this with the 
momentum quantization (\ref{mom_largeE2})}. This procedure is strongly analogous to the classification of the solutions of the KdV equations described 
above. These equations are hyperelliptic functions living on the same higher genus surface where the hyperelliptic curve is defined and parametrized by 
a \textit{set} of $\alpha$ s instead of one. We believe that it might be possible to generalize our results to the spiky case.

 \section{One loop correction to the energy}

Starting from the partition function and Wick rotating to the Euclidean time, the one-loop correction to the energy in the semiclassical limit;
 \bea
lim_{t\rightarrow\infty}\,Z=lim_{{\mathcal T}\rightarrow\infty}\,e^{-E_1\,\kappa\,{\mathcal T}} 
 \ea
can be obtained by integrating over the fluctuations, expanded up to the quadratic order in the static gauge. As usual, the resulting gaussian functional 
integrals will bring the determinants of the $2^{nd}$ order bosonic and fermionic fluctuation operators:
\bea
Z\approx \int{\mathcal D}\phi{\mathcal D}\beta_{1,2}{\mathcal D}\theta_{1,..,5}\,exp\left(-\frac{\sqrt{\lambda}}{4 \pi}\int d^2\sigma({\tilde {\mathcal L_B}}+
{\tilde {\mathcal L_F}})\right)=\sqrt{\frac{\det^4(\op_{\psi+})\det^4(\op_{\psi-})}{\det(\op_{\phi})\det^2(\op_{\beta})\det^5(\op_{\theta})}}
  \ea

As mentioned in the previous section, the two fermionic potentials $V_{\psi\pm}$ are self isospectral; thus their determinants are identical. For simplicity, 
from now on we will drop the $\pm$ sign as we work with the determinants. Writing $\ln\det\op$ as ${\rm tr}\ln\op$ the one loop energy is:
\bea
-E_1\,\kappa\,{\mathcal T}=\frac{1}{2}\rm{tr}_{\tau,\sigma} ln \frac{(-\dd-\partial^2_\tau+V_\psi(\sigma))^8}{(-\dd-\partial^2_\tau+V_\phi(\sigma))(-\dd-
\partial^2_\tau+V_\beta(\sigma))^2(-\dd-\partial^2_\tau)^5}
\ea
After Fourier transforming the $\tau$ coordinate,
 \bea
-\partial^2_\tau \rightarrow \Omega^2 \quad {\rm tr}_\tau\rightarrow{\mathcal T}\,\int \frac{d\Omega}{2\pi}
 \ea
the energy becomes the combination of various tr ln  terms:  
\bea 
E_1=-\frac{1}{2 \kappa}{\rm tr}_\sigma\,\int_{-\infty}^{\infty}\,\frac{d\Omega}{2 \pi} \ln \frac{(-\dd+V_\psi(\sigma)+\Omega^2)^8}{(-\dd+V_\phi(\sigma)+
\Omega^2)\,(-\dd+V_\beta(\sigma)+\Omega^2)^2\,(-\dd+\Omega^2)^5}
\ea
The $\Omega$ integral is convergent and can be evaluated:
\bea
E_1&=&\frac{1}{2 \kappa}{\rm tr}_\sigma\,\left( \sqrt{-\dd+V_\phi(\sigma)}+2\sqrt{-\dd+V_\beta(\sigma)}+5\sqrt{-\dd}-8\sqrt{-\dd+V_\psi(\sigma)} \right)\nn
&=&\frac{1}{2 \kappa}\sum_{eigenvalues}\left(\sqrt{\Ld_\phi}+2\sqrt{\Ld_\beta}+5n-8\sqrt{\Ld_\psi} \right)\nn
&=&\frac{1}{2 \kappa}\left[\sum_{i=I,II,III}\left(\sqrt{\Ld_{\phi,\,i}}+2\sqrt{\Ld_{\beta,\,i}}\right)+2(5-8\sqrt{\Ld_{\psi,\,0}}) \right. \nn &&\left.+2\sum_{n=1}^{\infty}\,
\left(\sqrt{\Ld_{\phi,\,n}}+2\sqrt{\Ld_{\beta,\,n}}+5(n+1)-8\sqrt{\Ld_{\psi,\,n}} \right)\right]
\label{esum}
\ea
where in the last line, the first sum is over the analytically known band edges, the second term is the $n=0$ term that contains only the $\theta$ and $\psi$ 
contributions.
As shown in the previous section, all the eigenvalues approach  the massless limit high up in the spectrum and the convergence can be seen through 
the cancellation of the bosonic degrees of freedom with the fermionic ones in the UV regime, illustrating the UV finiteness of the theory. The expression 
(\ref{esum}) can further be simplified by writing down the band edges (\ref{beta_edges},\,\ref{phi_edges}) explicitly. Recalling $\kappa=\frac{2\K}{\pi}k$ we 
get;
\bea
E_{BE}&=&\frac{1}{2 \kappa}\sum_{i=I,II,III}\left(\sqrt{\Ld_{\phi,\,i}}+2\sqrt{\Ld_{\beta,\,i}}\right)\nn&=&\frac{\pi}{4 \K k}\left(\frac{2\K }{\pi}(1-k)+\frac{2\K}{\pi}
(1+k)+2\frac{2\K }{\pi}k+2\frac{2\K }{\pi}+2\frac{2\K }{\pi}\sqrt{1+k^2}\right)\nn
&=&1+\frac{\sqrt{1+k^2}+2}{k}
\label{band_edges}
\ea 
In the rest of this section we evaluate the sum: 
\bea
\sum_{n=1}^{\infty}\,\left(\sqrt{\Ld_{\phi,\,n}}+2\sqrt{\Ld_{\beta,\,n}}+5(n+1)-8\sqrt{\Ld_{\psi,\,n}} \right)
\ea
by using Euler-Maclaurin formula and by converting the sum into a series of Riemann zeta functions.

\begin{figure}[h]
\center
\includegraphics[scale=0.6]{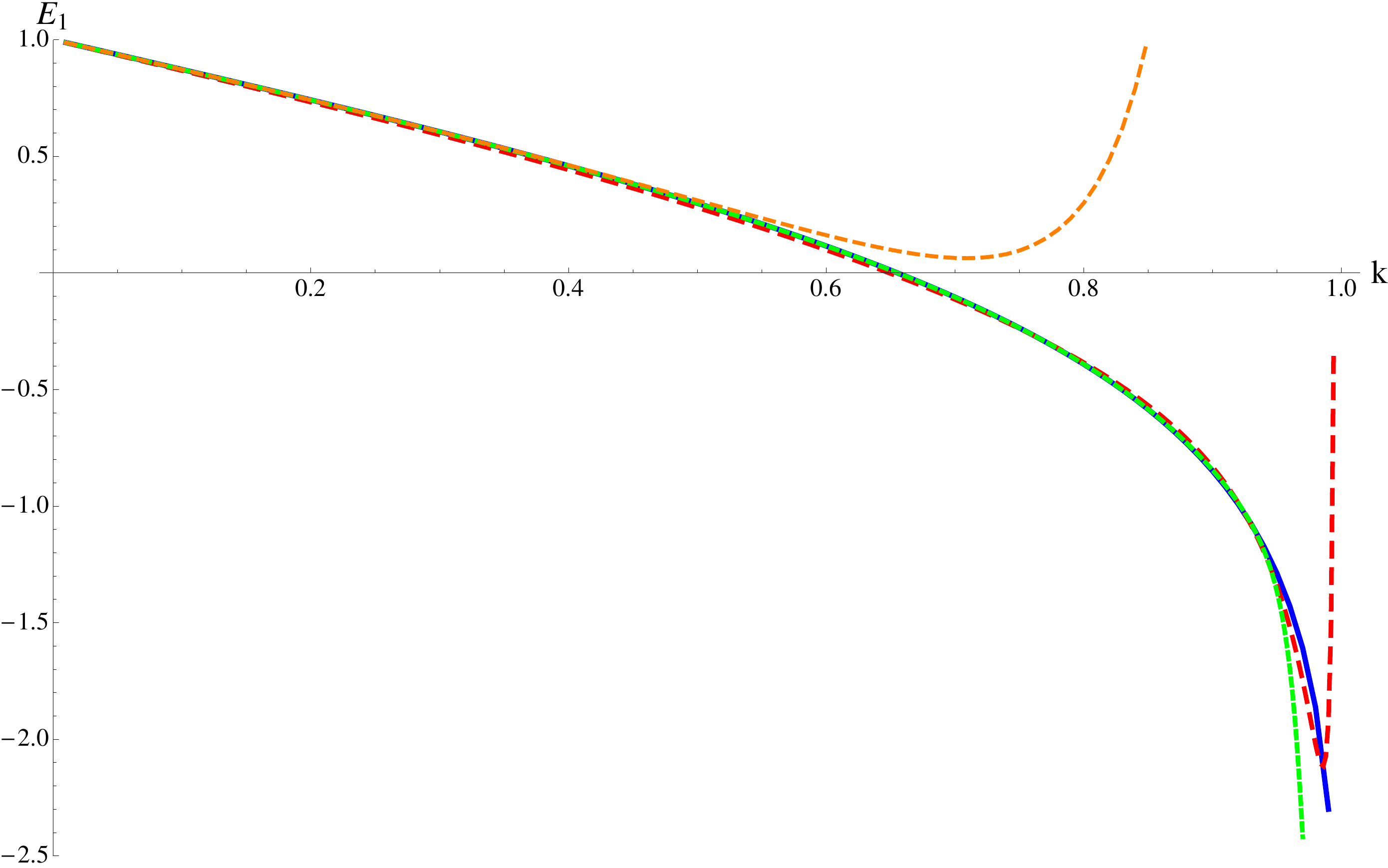}
\caption{The comparison of Euler-Maclaurin and zeta approaches to the small spin expression in \cite{Beccaria:2010ry}. Blue (solid), red(dashed) and 
green(dotted) are the results of numerical evaluation of the exact expression, Euler-Maclaurin approximation (\ref{EML}) and zeta function approximation (\ref{Ezeta}), 
respectively. The orange graph is the expression in \cite{Beccaria:2010ry}. Observe that  both of our results fits the exact expression perfectly except large 
spin regime. The zeta function approximation is better at small/medium $k$ regime but Euler-Maclaurin gets better as $k$ approaches 1.}
\label{energy-fig}
\end{figure}

Our results are plotted in figure \ref{energy-fig} as a function of the elliptic parameter $k$. Comparison with the exact numerical result from \cite{Beccaria:2010ry}, shows that they are extremely accurate in a considerably large region of $k$.  In particular, compared with the small $k$ expansion in \cite{Beccaria:2010ry}, both Euler-Maclaurin and zeta function approximations lead to a much better result. 

 \subsection{Euler-Maclaurin approach}
 In this part we expand the eigenvalues $\Ld_n$ up to order $(n+1)^{-2}$ using the results (\ref{beta_energy}), (\ref{phi_energy}) and (\ref{psi_energy}) for 
the $\beta$ , $\phi$ and fermion mode eigenvalues and approximate the sum by an integral by using the Euler-Maclaurin formula. Let us define:
 \bea
 f(n;k)&=&2\sqrt{(n+1)^2+2d_0+\frac{d_2}{(n+1)^2}}+\sqrt{(n+1)^2+4d_0+\frac{\td_2}{(n+1)^2}}\nn
 &&+5(n+1)-8\sqrt{(n+1)^2+d_0+\frac{d_2}{16(n+1)^2}}\nn
 \ea
The Euler-Maclaurin formula is given by:
 \bea
 \sum_{n=1}^{\infty}f(n;k)\approx\int_1^{\infty}f(n;k)dn+\frac{1}{2}f(1;k)-\sum_{j=1}^{M}\frac{B_{2j}}{(2j)!}f^{(2j-1)}(1;k)
 \label{f}
 \ea
 where $B_{2j}$s are the Bernoulli numbers ($B_2$=1/6, $B_4$=-1/30 etc..). We also used the fact that $f$ and all of its odd derivatives vanish as $n
\rightarrow\infty$. 

The integral in (\ref{f}) can be carried out analytically;
  \bea
I(k)&\equiv&\int_1^{\infty}f(n;k)dn=d_0\,\text{ln}\left[\frac{\left(4+d_0+\sqrt{16+8d_0+d_2}\right)\left(4+2 d_0+\sqrt{16+16 d_0+\td_2}\right)}{16\left(16+2 d_0+
\sqrt{64 (4+d_0)+\td_2}\right)^2}\right]\nn
&& -\sqrt{d_2}\, \text{ln}\left[\frac{\left(d_0+\sqrt{d_2}\right)\left(4 d_0+d_2+\sqrt{d_2 (16+8 d_0+d_2)}\right)}{d_2^3}\right]\nn
&&+\frac{\sqrt{\td_2}}{2}\,\text{ln}\left[\frac{16\left(2 d_0+\sqrt{\td_2}\right)\left(32 d_0+\td_2+\sqrt{\td_2 (64 (4+d_0)+\td_2)}\right)^2}{\td_2^3\left(8 d_0+
\td_2+\sqrt{\td_2 (16+16 d_0+\td_2)}\right)}\right]-f(1;k)
 \ea 
where the $I(k)$ is a function of $k$ through $d_0$, $d_2$ and $\td_2$ defined in section (\ref{spectral}). 
By taking the first correction $f(1;k)/2$ into account we obtain an explicit expression for the one loop energy as an explicit function of $k$:
\bea
E_1^{EML}(k)=1+\frac{\sqrt{1+k^2}+2}{k}+\frac{1}{\kappa(k)}(I(k)+\frac{1}{2}f(1;k)+5-8\Ld_{\psi\,0}(k))
\label{EML}
\ea
There is no simple expression for the lowest $\psi$ eigenvalue $\Ld_{\psi,\,0}$ but one can use the expansion (\ref{psi_energy}) and expand it up to any 
desired order. 

It is also straightforward to  extend the Euler-Maclaurin approach by including higher orders in the $(n + 1)^{-2}$ expansions  (\ref{beta_energy}), (\ref{phi_energy}), 
(\ref{psi_energy}) and evaluating the resulting integrals numerically. The advantage of restricting the order of the expansion to $(n+1)^{-2}$ is that simple 
analytic expressions can be given for the Euler-Maclaurin integrals.

  \subsection{Zeta function approach} 
In this section we follow another approach to calculate $E_1$. We start with expanding $\sqrt{\Ld_n}$ around large $n$ and and obtain a series in odd 
powers of $(n+1)^{-1}$. Then we sum over them and obtain a series in Riemann zeta functions evaluated at odd integers. The basic expansion is:
\bea
\sqrt{(n+1)^2+2d_0+\sum_{j=1}^{\infty}\frac{d_{2j}}{(n+1)^{2j}}}\approx(n+1)+\frac{d_0}{(n+1)}+\frac{d_2-d_0^2}{(n+1)^3}+ \ord((n+1)^{-5})
\label{n_exp}
\ea 
and including all the modes;
\bea
\sqrt{\Ld_{\phi,\,n}}+2\sqrt{\Ld_{\beta,\,n}}+5(n+1)-8\sqrt{\Ld_{\psi,\,n}}\approx\sum_{j=1}^{\infty}\frac{z_{2j+1}(k)}{(n+1)^{2j+1}}
\label{Ezeta_exp}
\ea
The $(n+1)$ and $1/(n+1)$ terms cancel out illustrating the UV finiteness once more. The coefficients are polynomials in $\EK$ and $k^2$ as 
expected:
\bea
z_3&=&16\left(\frac{\K}{\pi}\right)^4\left(-4\left(\EK\right)^2-(7+k^2)\EK+(-3+k^2) \right) \nn
z_5&=&\frac{16}{3}\left(\frac{\K}{\pi}\right)^6\left(-216\left(\EK\right)^3-8(-70+11k^2 ) \left(\EK\right)^2-7(481-192k^2+9k^4 )\EK \right. \nn &&+\left.(137-98k^2+15k^4))\right) 
\nn
\ea
We again begin with the summation:
\bea
\sum_{n=1}^{\infty}\,\left(\sqrt{\Ld_{\phi,\,n}}+2\sqrt{\Ld_{\beta,\,n}}+5(n+1)-8\sqrt{\Ld_{\psi,\,n}} \right)
\label{sum}
\ea
 It can be clearly seen that using the expansion (\ref{Ezeta_exp}) in the sum (\ref{sum}) leads to series of zeta functions with coefficients $z_n$.
 \bea
\sum_{n=1}^{\infty}\,\sum_{j=1}^{\infty}\frac{z_{2j+1}(k)}{(n+1)^{2j+1}}=\sum_{j=1}^{\infty} z_{2j+1}(k) (\zeta(2j+1)-1)
\ea
This is the zeta function analogue of the Euler-Maclaurin result (\ref{f}). The other terms in the expansion are exactly the same as the EM expression.
\bea
E_1^{\zeta}=1+\frac{\sqrt{1+k^2}+2}{k}+\frac{1}{\kappa(k)}\left(5-8\Ld_{\psi\,0}(k)+\sum_{j=1}^{\infty}\,z_{2j+1}(k)\,(\zeta(2j+1)-1)\right)
\label{Ezeta}
\ea

As seen from figure \ref{energy-fig}, the zeta function approach works well in the small/medium $k$ region, but breaks down as $k$ 
approaches very close to 1. The reason is that, in the expansion (\ref{n_exp}) of $\Ld_n$, we assumed the coefficients of $(n+1)^{-2j}$ are small; however 
their magnitudes grow with $k$, causing the expansion to fail near $k$ = 1. On the other hand, the Euler-Maclaurin approximation does not have that problem and it 
gets better than the zeta approach as $k$ increases, but it is limited to order $n^{-2}$. Of course, the WKB series of $\Ld_n$ (\ref{beta_energy},\,\ref{phi_energy},\,\ref{psi_energy}) gets worse around $k$=1 and both Euler-Maclaurin and zeta approximations eventually break down as $k$ gets extremely close to 1. Nevertheless, in the small and intermediate $k$ range, both the Euler-Maclaurin and zeta approaches lead to better approximations than the small $k$ expansion studied in \cite{Beccaria:2010ry}.

Some issues have been raised about the applicability of the zeta function expansion \cite{SchaferNameki:2005is}. The main problem there is that the zeta 
function approach fails to capture the leading term that Euler-Maclaurin integral captures. The source of the problem is, the zeta function approximation is 
valid when $n$ is large in the expression $\sqrt{n^2+\kappa^2}$ which is not the case in the large spin regime where the masses $\kappa$ are large. We do not encounter that problem here because we are outside of the $k\approx1$ regime.  We also do not take into account the 
exponentially suppressed corrections since they do not play a crucial role in the small/medium spin regime. 

Consequently, both the Euler-Maclaurin and zeta function approaches capture a remarkable portion of the one-loop energy as a function of the elliptic 
parameter and can be considered as an improvement of the previous result \cite{Beccaria:2010ry}. 

\section{Small spin expansion}

After expanding $\Sp(k)$ (\ref{classical_spin}) around $k=0$ and inverting  the series, we can express the elliptic parameter as a function of spin:
\bea
k^2\approx 2 \Sp -\frac{9}{2}\Sp^2 + \frac{87}{8}\Sp^3+\ord(\Sp^4)
\label{sk}
\ea
By plugging (\ref{sk}) into the zeta function expansion of the one-loop energy (\ref{Ezeta}), we find the one-loop energy as a series in $\sqrt{\Sp}$:
\bea
E_1&\approx&1+\sqrt{2\Sp}\left(\frac{1}{2^7} \left(-608+64\,\zeta(3)+112\,\zeta(5)+124\,\zeta(7)+127\,\zeta(9)\right)\right.\nn 
&&+\frac{\Sp}{2^{17}}(3352121-90112\,\zeta(3)-354304\,\zeta(5)-967168\,\zeta(7)-1836672\,\zeta(9))\nn 
&&+\left.\frac{\Sp^2}{2^{26}}(-4708374189+50069504\,\zeta(3)+241762304\,\zeta(5)+1019887616\,\zeta(7)\right. \nn 
&&+\left.3363704832\,\zeta(9))+\ord(\Sp^3)\right)\nn
\label{spin_zeta}
\ea
It is an asymptotic series with growing coefficients and the optimum order seems to be $\Sp^{5/2}$. Regardless of its asymptotic nature, the series 
(\ref{spin_zeta}) can be improved by taking more terms in the zeta function expansion (\ref{Ezeta}).\footnote{The fermion mode $\Ld_{\psi\,0}$ is expanded 
up to the same order with the zeta terms in the zeta function expansion.} By this way one generates more zeta functions in each coefficient. In other words, 
each coefficient in the short spin expansion can be expressed as infinite series of zeta functions. Furthermore, the coefficient of $\sqrt{2\Sp}$ follows a 
simple pattern which is $\sum_{j=1}^{\infty}\frac{2^{2j-1}-1}{2^{2j-1}}\zeta(2j+1)$. Remarkably it satisfies the identity:
 \bea
\sum_{j=1}^{\infty}\frac{2^{2j-1}-1}{2^{2j-1}}\zeta(2j+1)=\frac{3}{2}-4\,\text{ln}(2)
\ea
which is exactly the same factor in \cite{Beccaria:2010ry,short} derived in a different way, namely expanding the analytical expressions of the 
determinants in small spin limit. We could not found a simple pattern for the coefficients of the higher order terms. However, numerically they approach to 
the values in \cite{Beccaria:2010ry}. 

\begin{table}[h]
\center
\begin{tabular}{llcc}
coefficient of & actual value & up to order $\zeta(9)$& up to order $\zeta(17)$ \\
 \cline{1-4} \\ 
$\Sp^{1/2}$ & $\frac{3}{2}-4\text{ln}(2)$ =													-1.79971& 		 -1.80064 	 &			 -1.79971 \\ \\
$\Sp^{3/2}$ &$-\frac{23}{16} + \frac{3}{2} \text{ln}(2) + \frac{3}{4} \zeta(3)$=						 0.712429&	  	 0.736544  &			   0.712394  \\ \\
$\Sp^{5/2}$ &$\frac{689}{256} - \frac{63}{32}\text{ln}(2)- \frac{15}{32} \zeta(3) - \frac{15}{16} \zeta(5)$=		-0.295304 &		 - 0.468007&	          - 0.293531 
\end{tabular}
\label{short_spin_comp}
\caption{A comparison of the coefficients of the short spin expansion in \cite{Beccaria:2010ry} with the coefficients obtained by the zeta function 
expansions. Notice that the agreement gets better as more terms in the zeta function expansion are considered.}
\end{table}

\begin{figure}[H]
\center
\includegraphics[scale=0.59]{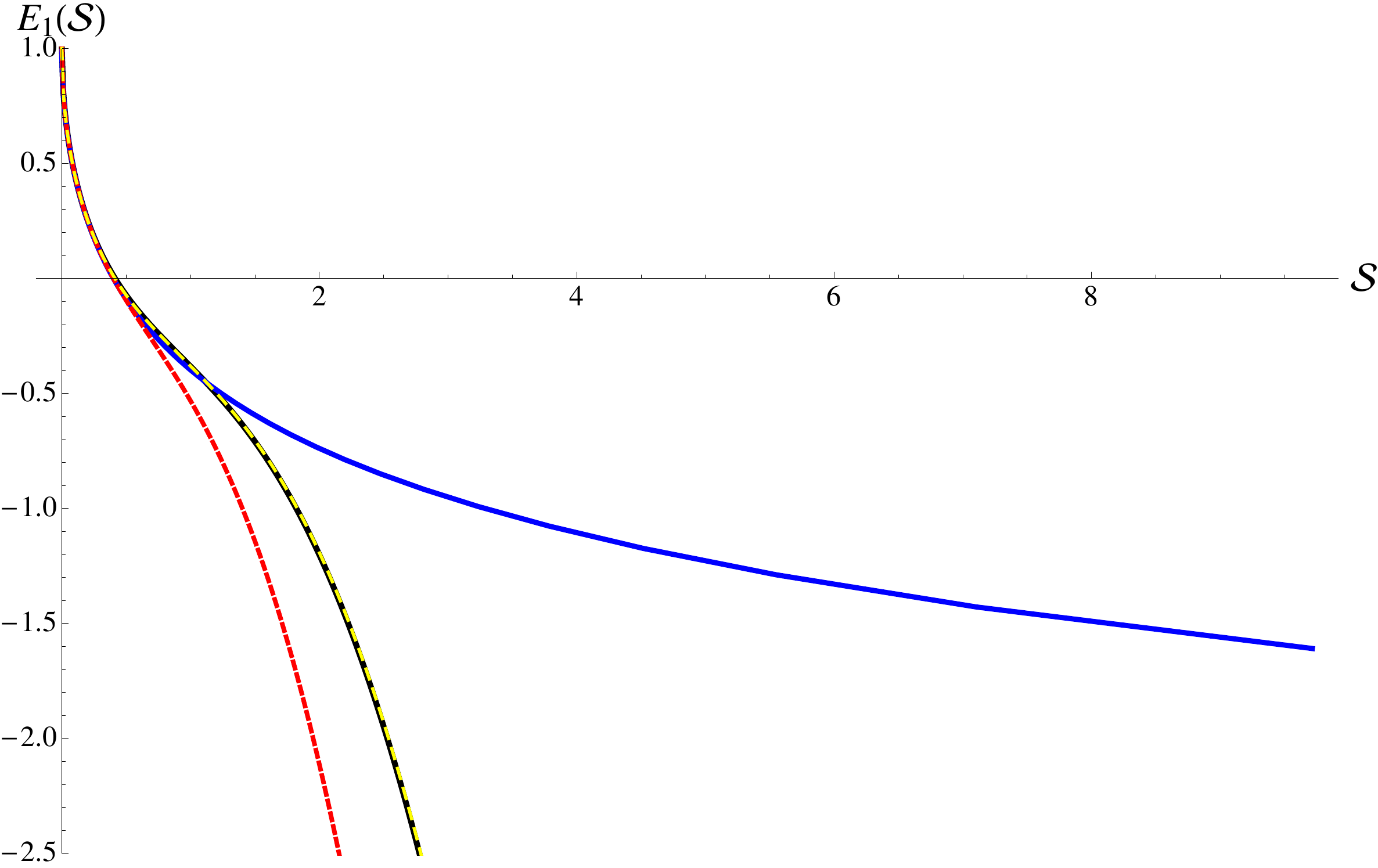}
\caption{The one-loop energy as a function of the spin $\Sp$. The blue (solid), red (dotted), yellow(dashed), black (solid) curves are the exact result, zeta 
expansion up to $\zeta(9)$ \ref{spin_zeta}, $\zeta(17)$ and the short spin expansion in \cite{Beccaria:2010ry}. }
\label{spin-fig}
\end{figure}

\section{Conclusion}

As a conclusion, we have constructed two different expansions for the one-loop energy of the folded spinning superstring in $AdS_5 \times S^5$ by summing over the fluctuation eigenvalues. 

The eigenvalues $\Lambda_n$ are calculated as series in inverse powers of $(n+1)$ up to any desired order in the semiclassical framework and two alternative mechanisms are introduced to construct the expansion. In more general grounds, it has been showed that the expansion is of the form: $\Ld_n=(n+1)^2+\langle V \rangle + \frac{-\langle V^2 \rangle+\langle V \rangle}{4\,(n+1)^2}+...$ where $V$ is the fluctuation potential induced by the target space curvature of the string sigma model. Since it is essentially a derivative expansion, the semiclassical approximation gets better in the flat space limit (i.e $\Sp\ll1$) where the potentials are slowly varying, yet is remarkably powerful in taking into account of the curvature corrections and extends to the limit where the end of the string is extremely close to the AdS boundary. It should be mentioned that this semiclassical framework is slightly different than the BMN \cite{Berenstein:2002jq} and FT \cite{Frolov:2002av} results where the quantum numbers ($S^5$ angular momentum $\mathcal{J}$ in the former, $\Sp$ in the latter) are large.

Then we used the eigenvalue expansion to calculate the one loop energy as a function of the elliptic parameter $k$. Again two alternative methods, namely Euler-Maclaurin and zeta function approximations, have been used. We found that the zeta function approximation is better for shorter strings but Euler-Maclaurin approximation takes over as the string gets longer. By comparing with the numerical result (see fig. \ref{energy-fig}), it is also demonstrated that for each of these approximation schemes, the final result is remarkably successful in capturing the exact expression in a wide range of the elliptic parameter. Finally we obtained the small spin Regge relation from the zeta function series and showed that (see fig \ref{spin-fig}) it leads to the correct result obtained by \cite{Beccaria:2010ry}. 

Our results may be helpful to resolve some issues of the strong vs weak coupling regimes of short operators. In contrast to the long operators that has the "universal" behavior $E=S+f(\lambda)\,$ln$S$+... for an arbitrary $\lambda$, the short operators have qualitatively different dependence on $S$: In the weak coupling limit, $E=g_1(\lambda)+g_2(\lambda)S+..$ where $g_i$ s are analytic functions of $\lambda$ and in the strong coupling limit  $E=1+\sqrt{S}(h_1(\lambda)+h_2(\lambda)S+..)$ where $h_i$ s have expansions in $\frac{1}{\sqrt{\lambda}}$. This suggests that somewhere in the parameter space ($S$,\,$\lambda$), the universal behavior of the scaling dimension is lost and the $S$ and $\lambda$ limits do not commute \cite{short}. Technically, it seems challenging to write the exact Regge relations even for the classical expression (\ref{classical_energy},\,\ref{classical_spin}). However the new way of obtaining the small $\Sp$ correction through the zeta function expansion may be useful to understand the behavior of the $\Sp$ expansion and ultimately lead to a better understanding of the nature of the $h_i$ coefficients. One can also make comparisons with the results on the general $\lambda$ dependence of the short operators presented in \cite{Gromov}. 

 Finally, it may possible to extend the heat kernel method presented in section \ref{heat_kernel} to coupled fluctuations in order to work out the eigenvalues for the configurations with both nonzero spin $S$ and R-charge $J$, where the fluctuations are neither decoupled nor soluble. Furthermore, this heat kernel expansion procedure can be compared with the integrability methods used in the spin chain framework.

\bigskip

I acknowledge support from the DOE  grant DE-FG02-92ER40716, 
and thank G. Dunne for discussions and comments.


\begin{thebibliography}{999}

\bibitem{Maldacena}
  J.~M.~Maldacena,
  ``The large N limit of superconformal field theories and supergravity,''
  Adv.\ Theor.\ Math.\ Phys.\  {\bf 2}, 231 (1998)
  [Int.\ J.\ Theor.\ Phys.\  {\bf 38}, 1113 (1999)]
  [arXiv:hep-th/9711200].

\bibitem{Gubser:2002tv}
  S.~S.~Gubser, I.~R.~Klebanov and A.~M.~Polyakov,
  ``A semi-classical limit of the gauge/string correspondence,''
  Nucl.\ Phys.\  B {\bf 636}, 99 (2002)
  [arXiv:hep-th/0204051].

\bibitem{Berenstein:2002jq}
  D.~E.~Berenstein, J.~M.~Maldacena and H.~S.~Nastase,
  ``Strings in flat space and pp waves from N = 4 super Yang Mills,''
  JHEP {\bf 0204}, 013 (2002)
  [arXiv:hep-th/0202021].

\bibitem{deVega:1996mv}
  H.~J.~de Vega and I.~L.~Egusquiza,
  ``Planetoid String Solutions in 3 + 1 Axisymmetric Spacetimes,''
  Phys.\ Rev.\  D {\bf 54}, 7513 (1996)
  [arXiv:hep-th/9607056].

\bibitem{Frolov:2002av}
  S.~Frolov and A.~A.~Tseytlin,
  ``Semiclassical quantization of rotating superstring in AdS(5) x S(5),''
  JHEP {\bf 0206}, 007 (2002)
  [arXiv:hep-th/0204226].

\bibitem{Beccaria:2010ry}
  M.~Beccaria, G.~V.~Dunne, V.~Forini, M.~Pawellek and A.~A.~Tseytlin,
  ``Exact computation of one-loop correction to energy of spinning folded
  string in $AdS_5 x S^5$,''
  J.\ Phys.\ A {\bf 43}, 165402 (2010)
  [arXiv:1001.4018 [hep-th]].




\bibitem{algcurve1}
  N.~Beisert and M.~Staudacher,
  ``Long-range PSU(2,2|4) Bethe ansaetze for gauge theory and strings,''
  Nucl.\ Phys.\  B {\bf 727}, 1 (2005)
  [arXiv:hep-th/0504190].
  N.~Beisert, V.~A.~Kazakov, K.~Sakai and K.~Zarembo,
  ``The algebraic curve of classical superstrings on $AdS(5) x S^5$,''
  Commun.\ Math.\ Phys.\  {\bf 263}, 659 (2006)
  [arXiv:hep-th/0502226].
  
    
 \bibitem{algcurve2}
  N.~Gromov,
  ``Integrability in AdS/CFT correspondence: Quasi-classical analysis,''
  J.\ Phys.\ A  {\bf 42}, 254004 (2009).
  N.~Gromov and P.~Vieira,
  ``The $AdS(5) x S^5$ superstring quantum spectrum from the algebraic curve,''
  Nucl.\ Phys.\  B {\bf 789}, 175 (2008)
  [arXiv:hep-th/0703191].




\bibitem{SchaferNameki:2005is}
  S.~Schafer-Nameki and M.~Zamaklar,
  ``Stringy sums and corrections to the quantum string Bethe ansatz,''
  JHEP {\bf 0510}, 044 (2005)
  [arXiv:hep-th/0509096].


  
\bibitem{integ}  
     S.~Schafer-Nameki, M.~Zamaklar and K.~Zarembo,
  ``Quantum corrections to spinning strings in AdS(5)xS(5) and Bethe ansatz: a
  comparative study,''
  JHEP {\bf 0509}, 051 (2005)
  [arXiv:hep-th/0507189].
 
  N.~Beisert and A.~A.~Tseytlin,
  ``On quantum corrections to spinning strings and Bethe equations,''
  Phys.\ Lett.\  B {\bf 629}, 102 (2005)
  [arXiv:hep-th/0509084].








\bibitem{4loop}
  Z.~Bern, M.~Czakon, L.~J.~Dixon, D.~A.~Kosower and V.~A.~Smirnov,
  ``The Four-Loop Planar Amplitude and Cusp Anomalous Dimension in Maximally 
  Supersymmetric Yang-Mills Theory,''
  Phys.\ Rev.\  D {\bf 75}, 085010 (2007)
  [arXiv:hep-th/0610248].
F.~Cachazo, M.~Spradlin and A.~Volovich,
  ``Four-Loop Cusp Anomalous Dimension From Obstructions,''
  Phys.\ Rev.\  D {\bf 75}, 105011 (2007)
  [arXiv:hep-th/0612309].
  T.~Lukowski, A.~Rej and V.~N.~Velizhanin,
  ``Five-Loop Anomalous Dimension of Twist-Two Operators,''
  Nucl.\ Phys.\  B {\bf 831}, 105 (2010)
  [arXiv:0912.1624 [hep-th]].


\bibitem{novel}
  N.~Beisert, V.~Dippel and M.~Staudacher,
  ``A novel long range spin chain and planar N = 4 super Yang-Mills,''
  JHEP {\bf 0407}, 075 (2004)
  [arXiv:hep-th/0405001].

\bibitem{Gromov}
  N.~Gromov, V.~Kazakov and P.~Vieira,
  ``Exact Spectrum of Planar ${\cal N}=4$ Supersymmetric Yang-Mills Theory:
  Konishi Dimension at Any Coupling,''
  Phys.\ Rev.\ Lett.\  {\bf 104}, 211601 (2010)
  [arXiv:0906.4240 [hep-th]].
  N.~Gromov, V.~Kazakov, A.~Kozak and P.~Vieira,
  ``Exact Spectrum of Anomalous Dimensions of Planar N = 4 Supersymmetric
  Yang-Mills Theory: TBA and excited states,''
  Lett.\ Math.\ Phys.\  {\bf 91}, 265 (2010)
  [arXiv:0902.4458 [hep-th]].
  N.~Gromov, V.~Kazakov and P.~Vieira,
  ``Exact Spectrum of Anomalous Dimensions of Planar N=4 Supersymmetric
  Yang-Mills Theory,''
  Phys.\ Rev.\ Lett.\  {\bf 103}, 131601 (2009)
  [arXiv:0901.3753 [hep-th]].




\bibitem{dorey}
  N.~Dorey,
  ``A Spin Chain from String Theory,''
  Acta Phys.\ Polon.\  B {\bf 39}, 3081 (2008)
  [arXiv:0805.4387 [hep-th]].
   N.~Dorey and M.~Losi,
  ``Spiky Strings and Spin Chains,''
  arXiv:0812.1704 [hep-th].



\bibitem{spiky}
    M.~Kruczenski and A.~Tirziu,
  ``Spiky strings in Bethe Ansatz at strong coupling,''
  Phys.\ Rev.\  D {\bf 81}, 106004 (2010)
  [arXiv:1002.4843 [hep-th]].
  L.~Freyhult, M.~Kruczenski and A.~Tirziu,
  ``Spiky strings in the SL(2) Bethe Ansatz,''
  JHEP {\bf 0907}, 038 (2009)
  [arXiv:0905.3536 [hep-th]].



  \bibitem{bethe}
   N.~Beisert, B.~Eden and M.~Staudacher,
  ``Transcendentality and crossing,''
  J.\ Stat.\ Mech.\  {\bf 0701}, P021 (2007)
  [arXiv:hep-th/0610251].
    B.~Eden and M.~Staudacher,
  ``Integrability and transcendentality,''
  J.\ Stat.\ Mech.\  {\bf 0611}, P014 (2006)
  [arXiv:hep-th/0603157].


\bibitem{belitsky}
   A.~V.~Belitsky, V.~M.~Braun, A.~S.~Gorsky and G.~P.~Korchemsky,
  ``Integrability in QCD and beyond,''
  Int.\ J.\ Mod.\ Phys.\  A {\bf 19}, 4715 (2004)
  [arXiv:hep-th/0407232].
    A.~V.~Belitsky, A.~S.~Gorsky and G.~P.~Korchemsky,
  ``Logarithmic scaling in gauge / string correspondence,''
  Nucl.\ Phys.\  B {\bf 748}, 24 (2006)
  [arXiv:hep-th/0601112].



  
















\bibitem{Benna:2006nd}
  M.~K.~Benna, S.~Benvenuti, I.~R.~Klebanov and A.~Scardicchio,
  ``A test of the AdS/CFT correspondence using high-spin operators,''
  Phys.\ Rev.\ Lett.\  {\bf 98}, 131603 (2007)
  [arXiv:hep-th/0611135].







\bibitem{cusp}
  A.~M.~Polyakov,
  ``Gauge Fields As Rings Of Glue,''
  Nucl.\ Phys.\  B {\bf 164} (1980) 171.
  G.~P.~Korchemsky,
  ``Asymptotics of the Altarelli-Parisi-Lipatov Evolution Kernels of Parton
  Distributions,''
  Mod.\ Phys.\ Lett.\  A {\bf 4}, 1257 (1989).
  
  \bibitem{QCD}
  D.~J.~Gross and F.~Wilczek,
  ``Asymptotically Free Gauge Theories. 2,''
  Phys.\ Rev.\  D {\bf 9}, 980 (1974).
  H.~Georgi and H.~D.~Politzer,
  ``Electroproduction scaling in an asymptotically free theory of strong
  interactions,''
  Phys.\ Rev.\  D {\bf 9}, 416 (1974).
  








  

\bibitem{Basso:2007wd}
  B.~Basso, G.~P.~Korchemsky and J.~Kotanski,
  ``Cusp anomalous dimension in maximally supersymmetric Yang-Mills theory at
  strong coupling,''
  Phys.\ Rev.\ Lett.\  {\bf 100}, 091601 (2008)
  [arXiv:0708.3933 [hep-th]].






\bibitem{long}
  S.~Frolov, A.~Tirziu and A.~A.~Tseytlin,
  ``Logarithmic corrections to higher twist scaling at strong coupling from
  AdS/CFT,''
  Nucl.\ Phys.\  B {\bf 766}, 232 (2007)
  [arXiv:hep-th/0611269].
  M.~Beccaria, V.~Forini, A.~Tirziu and A.~A.~Tseytlin,
  ``Structure of large spin expansion of anomalous dimensions at strong
  coupling,''
  Nucl.\ Phys.\  B {\bf 812}, 144 (2009)
  [arXiv:0809.5234 [hep-th]].

\bibitem{short}
  A.~Tirziu and A.~A.~Tseytlin,
  ``Quantum corrections to energy of short spinning string in AdS5,''
  Phys.\ Rev.\  D {\bf 78}, 066002 (2008)
  [arXiv:0806.4758 [hep-th]].
  
   R.~Roiban and A.~A.~Tseytlin,
  ``Quantum strings in $AdS_5 x S^5$: strong-coupling corrections to dimension of  Konishi operator,''
  JHEP {\bf 0911}, 013 (2009)
  [arXiv:0906.4294 [hep-th]].




 








\bibitem{Dunne:1999uy}
  G.~V.~Dunne and T.~M.~Hall,
  ``Borel summation of the derivative expansion and effective actions,''
  Phys.\ Rev.\  D {\bf 60}, 065002 (1999)
  [arXiv:hep-th/9902064].

\bibitem{Correa:2009xa}
  F.~Correa, G.~V.~Dunne and M.~S.~Plyushchay,
  ``The Bogoliubov/de Gennes system, the AKNS hierarchy, and nonlinear quantum
  mechanical supersymmetry,''
  Annals Phys.\  {\bf 324}, 2522 (2009)
  [arXiv:0904.2768 [hep-th]].




\bibitem{Dunne:1997ia}
  G.~V.~Dunne and J.~Feinberg,
  ``Self-isospectral periodic potentials and supersymmetric quantum
  mechanics,''
  Phys.\ Rev.\  D {\bf 57}, 1271 (1998)
  [arXiv:hep-th/9706012].






 



\bibitem{magnus}
  W. Magnus and S. Winkler, {\it HillÕs Equation}, (Wiley, New York, 1966).

\bibitem{whittaker}
 E. T. Whittaker, G. N. Watson, {\it A course of modern analysis}, Cambridge University Press, 4th edition (1927).

\bibitem{algebro}
E. D. Belokos, A. I. Bobenko, V. Z. Enolskii, A. R. Its, V. B. Matveev, {\it Algebro-geometric approach to nonlinear integrable equations}, Springer Series in Nonlinear Dynamics, 
Springer, Berlin (1994). F. Gesztesy, H. Holden, J. Michor, and G. Teschl: {\it Soliton Equations and Their Algebro-Geometric Solutions}, Cambridge studies in advanced mathematics, volume 114 Cambridge University Press, Cambridge, 2008 

\bibitem{abram}
Abramowitz, Milton; Stegun, Irene A., eds. , {\it Handbook of Mathematical Functions with Formulas, Graphs, and Mathematical Tables}, New York: Dover Publications, 1972



\end{thebibliography}
\end{document}